\documentstyle[prd,aps,preprint,tighten,eqsecnum]{revtex}
\begin{document}
\preprint{gr-qc/9506085} 
\title{Black Hole Entropy and the Hamiltonian Formulation of\\
          Diffeomorphism Invariant Theories}
\author{J. David Brown}
\address{Department of Physics and Department of Mathematics,\\
   North Carolina State University, Raleigh, NC 27695--8202}
\maketitle
\begin{abstract}
Path integral methods are used to derive a general expression for
the entropy of a black hole in a diffeomorphism invariant theory.
The result, which depends on the variational derivative
of the Lagrangian with respect to the Riemann tensor, agrees
with the result obtained {}from Noether charge methods by Iyer and Wald.
The method used here is based on the direct expression
of the density of states as a path integral (the microcanonical
functional integral). The analysis makes crucial use of the
Hamiltonian form of the action. An algorithm for placing the
action of a diffeomorphism invariant theory in Hamiltonian form is
presented. Other path integral approaches to the derivation of
black hole entropy include the Hilbert
action surface term method and the conical deficit angle method.
The relationships between these path integral methods
are presented.
\end{abstract}
\pacs{???}
\section{Introduction}
Noether charge methods have led Iyer and Wald \cite{Wald,IW1} to the
discovery of two elegant expressions for the entropy of a stationary
black hole in a diffeomorphism invariant theory in $D$
spacetime dimensions. The first expression is
\begin{equation}
{\cal S}_{\scriptscriptstyle BH} = 2\pi \int_{{\cal H}} {\bbox{Q}}[t] \
,\eqnum{1.1}
\end{equation}
where ${\cal H}$ denotes the black hole bifurcation surface and
${\bbox{Q}}[t]$
is the Noether charge ($D-2$)--form associated with the horizon Killing
field $t^a$. The second expression is
\begin{equation}
{\cal S}_{\scriptscriptstyle BH} = -2\pi \int_{{\cal H}} d^{D-2}x
\sqrt{\sigma}
\epsilon_{ab}\epsilon_{cd} U_0^{abcd}
 \ ,\eqnum{1.2}
\end{equation}
where ${\sigma}$ is the determinant of the metric on ${\cal H}$,
$\epsilon_{ab}$ is the binormal of ${\cal H}$, and
$U_0^{abcd}$ is the variational derivative \cite{Donder}
of the Lagrangian with respect to the Riemann tensor
${\cal R}_{abcd}$. Equation (1.2) is a generalization of the result
obtained in Ref.~\cite{JKM} for black hole entropy in a theory
described by a Lagrangian that depends on at most first derivatives
of the Riemann tensor. The equivalence of expressions (1.1) and
(1.2) is demonstrated in Ref.~\cite{IW1}.

More recently, Iyer and Wald \cite{IW2} and Nelson \cite{Nelson}
compared the Noether charge approach with various path integral
derivations of black hole entropy. The path integral methods all
originate, ultimately, with the observation made by Gibbons and
Hawking \cite{GH} that the partition function for the gravitational
field can be expressed as a path integral.\footnote{The original
calculation of Gibbons and Hawking was inconsistent. Their result
for the partition function implies a negative value for the heat
capacity. On the other hand, general arguments show that the heat
capacity is necessarily positive for any system that can be
characterized by a partition function. This problem was overcome
by York \cite{York86} who showed that the partition function yields
a positive value for the heat capacity if the
boundary conditions in the path integral are imposed at a finite
spatial location.} These path integral methods were
developed within the context of specific theories, such as Einstein
gravity
or Lovelock gravity. They include (i) the direct expression of
$\exp({\cal S}_{\scriptscriptstyle BH})$ in terms of a path integral (the
microcanonical
functional integral) \cite{BY1,BY2}; (ii) the expression of
${\cal S}_{\scriptscriptstyle BH}$ in terms of the Hilbert action surface
term \cite{KOP,BTZ};
and (iii) the derivation of ${\cal S}_{\scriptscriptstyle BH}$ in terms of
a nonclassical
spacetime with a conical singularity \cite{BTZ,SU}.
Using the language and techniques of the Noether charge formalism, Iyer
and Wald showed that the path integral methods (i) and (ii)
yield the result (1.1), the black hole entropy expressed as the integral
of
the Noether charge ${\bbox{Q}}[t]$, when applied to an arbitrary
diffeomorphism invariant theory. Nelson has analyzed the relationship
between the path integral method (iii) and the Noether charge result
(1.1).

Section 4 of this paper contains a direct derivation of the result (1.2),
the black hole entropy expressed in terms of the variational derivative
of the Lagrangian. This derivation is based on the microcanonical functional
integral method (i) and bypasses the Noether charge formalism altogether.
There may be some advantage to this. The Noether charge formalism is
a useful tool for deriving the first law of black hole mechanics but,
by itself, it does not provide a logically complete derivation
of black hole entropy. In order to extract the black hole
entropy {}from the first law of black hole mechanics one must, in principle,
supplement the Noether charge analysis with the quantum field theory
scattering calculation \cite{Hawking} that leads to the identification of
surface gravity (divided by $2\pi$) with black hole
temperature. On the other
hand, the path integral approach, although formal, does provide a logically
complete framework in which black hole entropy can be derived and
analyzed. For this reason, insights into the mysteries of black hole
entropy, such as its statistical origin, should be obtained more
easily {}from within the path integral formalism.

The microcanonical functional integral method (i)
is reviewed in Sec.~2. The derivation of Eq.~(1.2) in Sec.~4 makes
crucial use of the Hamiltonian form of the action, which is derived in
Sec.~3. The three path integral methods mentioned above, having a common
origin, are closely related to one another. The logical connections
between the microcanonical functional integral method (i), the Hilbert
action surface term method (ii), and the conical deficit angle method
(iii) are discussed in Sec.~5.

An important part of the present analysis is contained in Sec.~3, where
an algorithm is developed that allows one to place the action for a
diffeomorphism invariant theory in Hamiltonian form. Specifically,
it is shown that any action can be placed in ``almost Hamiltonian" form,
which differs {}from a true Hamiltonian form by the presence of
extra undifferentiated variables (referred to as the $\chi$'s) in the
Hamiltonian constraint ${\cal C}_\perp$. A true Hamiltonian form of the
action
is obtained when the $\chi$'s are eliminated through the solution of
their algebraic equations of motion. If the rank of the
matrix formed {}from the second derivatives of ${\cal C}_\perp$
with respect to the $\chi$'s is not maximal, then there are
constraints on the canonical variables. These constraints must be
added to the action via Lagrange multipliers. In practice, it might
not be possible to solve analytically the algebraic equations of motion
for the $\chi$'s, for example, if the equations of motion include
high--order polynomials. (It might also occur that the solution
of the equations of motion for the $\chi$'s is not unique. In that
case, the action and the system it describes splits into separate
self--consistent Hamiltonian theories.)
In Appendix A the familiar Hamiltonian form of the action for
Einstein gravity coupled to Maxwell electrodynamics is
derived using the algorithm developed in Sec.~3.

A fourth path integral method considered by Iyer and Wald is (iv) the
calculation of $\exp({\cal S}_{\scriptscriptstyle BH})$ as the enhancement
factor for the
rate of black hole pair creation relative to the pair creation rate for
matter distributions. They show \cite{IW2} that method (iv)
yields the entropy expression (1.1) when applied to an arbitrary
diffeomorphism invariant theory. In Ref.~\cite{Brown}, it was shown
that the enhancement in the black hole creation rate, method (iv),
must agree with the entropy as calculated {}from the microcanonical
functional
integral, method (i).  Therefore the results of
Ref.~\cite{Brown} along with those obtained here constitute a derivation
of Eq.~(1.2) as the enhancement factor for black hole pair creation
in a diffeomorphism invariant theory.

The analysis presented here applies to any stationary spacetime with
bifurcate Killing horizon. This includes not only certain black hole
spacetimes but also, for example, Rindler spacetime. With periodic
identifications in the extra dimensions, the bifurcation surface ${\cal H}$
of the Rindler horizon has the topology of a ($D-2$)--torus.
Equation~(1.2) gives the associated entropy. For definiteness,
I will typically use the terminology appropriate
for black hole spacetimes. Some key results concerning the surface
gravity of a bifurcate Killing horizon, which are used in the analysis
of Sec.~4, are derived in Appendix B.
\section{Microcanonical Functional Integral}
In the microcanonical functional integral formalism, the density of states
$\nu$ is expressed directly as a path integral \cite{BY1,BY2}:
\begin{equation}
\nu = \sum_{{\cal M}}\int Dg\,D\psi \,\exp\Bigl({{\cal S}[g,\psi]}\Bigr) \
.\eqnum{2.1}
\end{equation}
Here, ${\cal S}$ is the action, which is a
functional of the metric $g_{ab}$ and a collection of matter
fields denoted by $\psi$. Also, $\sum_{{\cal M}}$ denotes a sum over
manifolds
${\cal M}$ of different topologies, subject to the requirement that the
boundary $\partial{\cal M}$ should have topology ${\cal B}\times S^1$. For
the purpose
of describing the thermodynamics associated with a horizon, it is most
convenient to choose ${\cal B}$ to have the same topology as the
bifurcation
surface. Thus, for the case of black hole spacetimes, ${\cal B}$ is a
($D-2$)--sphere. For the case of Rindler spacetime ${\cal B}$ is a
($D-2$)--torus.

The boundary conditions on the metric and matter fields in the path
integral for the density of states $\nu$ involve fixation of those
quantities on $\partial{\cal M}$
that characterize the states of the system. In the terminology
of traditional thermodynamics, these are the extensive variables
including, for example, internal energy and electric charge. These
quantities appear at the classical level as functions of the canonical
variables $q^\alpha$ and $p_\alpha$. Thus, consider the action ${\cal S}$
written
in Hamiltonian form,
\begin{equation}
{\cal S}[\lambda,q,p] = i\int_{S^1} dt \int_\Sigma d^dx \Bigl( p_\alpha
{\dot q}^\alpha - \lambda^{\scriptscriptstyle A} {\cal
C}_{\scriptscriptstyle A}(q,p) \Bigr) +
(\hbox{boundary terms}) \ , \eqnum{2.2}
\end{equation}
where $d=D-1$ is the dimensionality of space $\Sigma$. The Lagrange
multipliers are denoted by $\lambda^{\scriptscriptstyle A}$ and ${\cal
C}_{\scriptscriptstyle A}(q,p)$ are
the constraints. The particular form (2.2) for the action follows {}from
spacetime diffeomorphism invariance and the assumption that under
reparametrizations in $t$ the canonical variables transform as scalars
and the Lagrange multipliers transform as scalar densities \cite{HT}.

There are two types of boundary terms that appear in the Hamiltonian
form of the action for a diffeomorphism invariant theory on a
manifold ${\cal M}$. The first is a term on the boundary
$\partial{\cal M}={\cal B}\times S^1$ of the spacetime manifold.
By the argument given in Ref.~\cite{Brown}, the action (2.2) appropriate
for
the density of states $\nu$  contains no such boundary terms at
$\partial{\cal M}$. Otherwise, if boundary terms at $\partial{\cal M}$
were present,
the boundary conditions would include fixation of quantities that depend
on the Lagrange multipliers (and, hence, do not depend solely on the
canonical variables $q^\alpha$ and $p_\alpha$).

The second type of boundary term that appears in the Hamiltonian
form of the action arises only if the boundary of space
$\Sigma$ includes an element ${\cal H}$ in addition to the generic leaf
${\cal B}$ of the foliation of $\partial{\cal M}={\cal B}\times S^1$; that
is, if
$\partial\Sigma = {\cal H} \cup{\cal B}$. This situation occurs in
particular when
the spacetime manifold has topology ${\cal M} = {\cal B}\times
\hbox{$I$\kern-3.8pt $R$}^2$ and the
leaves of the foliation terminate at a common surface ${\cal H}$,
considered to
be the ``origin" of the $\hbox{$I$\kern-3.8pt $R$}^2$ plane. Such boundary
terms are derived
as follows. Start with the action in Lagrangian form, expressed as an
integral over ${\cal M}$ (plus possible boundary terms). Now excise a
region
{}from ${\cal M}$ surrounding ${\cal H}$, so that ${\cal M}$ has the
product topology
${\cal B}\times({\rm annulus}) = \Sigma\times S^1$ (where $\Sigma ={\cal
B}\times I$,
with $I$ a real line interval). The boundary $\partial{\cal M}$ then
consists of
two copies of ${\cal B}\times S^1$, where one copy coincides with the
original
boundary of ${\cal M}$ and the other copy coincides with the boundary of
the
excised region. The passage {}from the Lagrangian form of the action to
the Hamiltonian form of the action proceeds as usual, with various
boundary terms appearing at the boundary of the excised region. One
then takes the limit in which the excised region shrinks to zero,
being careful to insure that the geometry is smooth at ${\cal H}$. The
second type of boundary term is a term on the boundary of the excised
region that survives this limit.

The entropy of a stationary black hole is computed as follows
\cite{BY1,BY2}.
First, express the Lorentzian solution in stationary coordinates,
$ds^2 = {\tilde g}_{ab}\, dx^a dx^b$, ${\bbox{\psi}}={\tilde \psi}$,
where ${\tilde g}_{ab}$ and ${\tilde \psi}$ are $t$ independent. Next,
choose boundary conditions for the path integral (2.1) that coincide with
the boundary values (as constructed {}from the canonical data of a
$t={\hbox{const}}$ slice with boundary element ${\cal B}$) of the black
hole
spacetime. The path integral for $\nu$ will have an extremum in the
topological sector ${\cal M}={\cal B}\times \hbox{$I$\kern-3.8pt $R$}^2$
that consists of the
complex black hole solution
$ds^2 = {\bar g}_{ab} \, dx^a dx^b$, ${\bbox{\psi}}={\bar \psi}$.
The complex black hole is obtained {}from the Lorentzian black hole
by the substitution $t\to -it$. The $t={\hbox{const}}$ slices of the
Lorentzian and complex black hole solutions coincide in the sense
that their canonical data agree \cite{BMY,BY1,Brown}. In particular,
the data on the boundary element ${\cal H}$ of the spatial slices of the
complex black hole coincide with the data on the bifurcation surface
of the Lorentzian
black hole. In the zero--loop approximation the density of states
is given by $\nu \approx \exp({\cal S}[{\bar g},{\bar\psi}])$, where
${\cal S}[{\bar g},{\bar\psi}]$ is the action evaluated at the complex
black hole solution. The entropy of the black hole is then
\begin{equation}
{\cal S}_{\scriptscriptstyle BH} \approx {\cal S}[{\bar g},{\bar\psi}] \
,\eqnum{2.3}
\end{equation}
the logarithm of the density of states.

When the action is expressed in Hamiltonian form, the evaluation
of the entropy in
Eq.~(2.3) is simple. Because the complex spacetime ${\bar g}$,
${\bar\psi}$ is a stationary solution of the classical equations of
motion, both the $p_\alpha{\dot q}^\alpha$ terms and the constraint
terms in Eq.~(2.2) vanish. The only contribution to the entropy comes
{}from the boundary terms. As discussed above (see Ref.~\cite{Brown}),
there are no boundary terms at $\partial{\cal M}$. There are, however,
boundary terms at ${\cal H}$. The entropy arises entirely
{}from the evaluation of these terms at the complex black hole solution.
\section{Action for Diffeomorphism Invariant Theories}
The action for an arbitrary diffeomorphism invariant theory of the
metric $g_{ab}$ and tensor matter fields $\psi$ can be expressed in the
manifestly covariant form \cite{IW1}
\begin{eqnarray}
& &{\cal S}[g,\psi] = i\int_{{\cal M}} d^Dx \sqrt{-g} \,{\cal L} \
,\eqnum{3.1a}\\
& &{\cal L} = {\cal L}\Bigl( g_{ab}, {\cal R}_{bcde},
\nabla_{a_1}{\cal R}_{bcde}, \ldots,
\nabla_{{\scriptscriptstyle (}a_1}\!\!\cdot\!\cdot\!\cdot\!
\nabla_{a_m{\scriptscriptstyle )}}{\cal R}_{bcde}, \psi,
\nabla_{a_1}\psi, \ldots,
\nabla_{{\scriptscriptstyle
(}a_1}\!\!\cdot\!\cdot\!\cdot\!\nabla_{a_\ell{\scriptscriptstyle )}}\psi
\Bigr)
\ ,\eqnum{3.1b}
\end{eqnarray}
where $\nabla_a$ is the spacetime covariant derivative.\footnote{The
work of Anderson and Torre \cite{AT} implies that
the Lagrangian (3.1b) can be written in terms of covariant derivatives of
the Riemann tensor in which the symmetrization over covariant derivatives
is extended to the second and fourth slots of the Riemann tensor itself.
The arguments of this section remain valid whether or not the Lagrangian
is written in this way.}
I will use the action (3.1) as the starting point. Note, however, that
the derivation of the entropy (1.2) in this paper is not restricted
just to the case of tensor matter fields.  For example, $\psi$ can include
the components $A^\alpha_a$ of the Yang--Mills connection ($\alpha$ is
the internal index), where the covariant derivative $\nabla_a$ of
Eq.~(3.1b) acts on $A^\alpha_a$ as a collection of covariant vectors.
Likewise, $\psi$ can include the tetrad field $(e^\mu)_a$. In this
case, the Lagrangian should include a
term $\Lambda^{ab}[ g_{ab} - (e^\mu)_a \eta_{\mu\nu} (e^\nu)_b]$ that
links the tetrad to the metric, where $\Lambda^{ab}$ is an independently
varied field (included among the $\psi$) and $\eta_{\mu\nu} = {\rm
diag}(-1,+1,\ldots,+1)$. With the tetrad appearing as a dynamical
variable one can also include coupling to the Dirac field (see,
for example, Ref.~\cite{NT}).

The goal of this section is to place the action (3.1) in Hamiltonian
form (2.2), and thereby derive the relevant boundary terms at ${\cal H}$.
I will assume that the manifold topology is ${\cal M}=\Sigma\times I$.
In Sec.~4, where the density of states (2.1) is evaluated,
the factor $I$ is periodically identified to form a circle $S^1$.
\subsection{Elimination of derivatives of the Riemann tensor}
The first step in the derivation of the Hamiltonian form of the action
(3.1) is the elimination of derivatives of the Riemann tensor. The
highest derivative, namely the $m^{\rm th}$ derivative, can be eliminated
as follows. Introduce  a set of auxiliary fields
$U_m^{a_1\cdots a_m bcde}$ and $V^m_{a_1\cdots a_m bcde}$, and write the
action as
\begin{eqnarray}
{\cal S}[g,\psi,U_m,V^m] & = & i\int_{{\cal M}} d^Dx \sqrt{-g} \biggl\{
{\cal L}\Bigl( g, {\cal R}, \nabla_{a_1}{\cal R}, \ldots,
\nabla_{{\scriptscriptstyle
(}a_1}\!\!\cdot\!\cdot\!\cdot\!\nabla_{a_{m-1}{\scriptscriptstyle )}}{\cal
R}, V^m_{a_1\cdots a_m},
{\hbox{$\psi$'s}} \Bigr) \nonumber\\
& &\qquad\qquad\qquad{\ } + U_m^{a_1\cdots a_m} \Bigl[
\nabla_{{\scriptscriptstyle (}a_1}
\!\!\cdot\!\cdot\!\cdot\!\nabla_{a_m{\scriptscriptstyle )}}{\cal R} \,-\,
V^m_{a_1\cdots a_m} \Bigr]
 \biggr\} \ .\eqnum{3.2}
\end{eqnarray}
Note that the indices on the Riemann tensor, and the corresponding
indices on $U_m$ and $V^m$, have been suppressed. Also, the notation
$\psi$'s is used for the matter fields and their derivatives. The
actions (3.1) and (3.2) are equivalent. This is demonstrated by
substituting the solution of the classical equations of motion for
$U_m$ and $V^m$, namely,
\begin{eqnarray}
0 & = & \frac{1}{i\sqrt{-g}} \frac{\delta{\cal S}}{\delta U_m^{a_1\cdots
a_m}}
= \nabla_{{\scriptscriptstyle
(}a_1}\!\!\cdot\!\cdot\!\cdot\!\nabla_{a_m{\scriptscriptstyle )}}{\cal R}
- V^m_{a_1\cdots a_m}
\ ,\eqnum{3.3a}\\
0 & = & \frac{1}{i\sqrt{-g}} \frac{\delta{\cal S}}{\delta V^m_{a_1\cdots
a_m}}
= -U_m^{a_1\cdots a_m} + \frac{\partial{\cal L}}{\partial V^m_{a_1\cdots
a_m}}
\ ,\eqnum{3.3b}
\end{eqnarray}
into the action (3.2). The result is the action (3.1). Now integrate
by parts in Eq.~(3.2) to remove one derivative {}from
$\nabla_{{\scriptscriptstyle
(}a_1}\!\!\cdot\!\cdot\!\cdot\!\nabla_{a_m{\scriptscriptstyle )}}{\cal
R}$, and discard the
boundary term. This leads to the action
\begin{eqnarray}
& & {\cal S}[g,\psi,U_m,V^m] \nonumber\\
& &\qquad = i\int_{{\cal M}} d^Dx \sqrt{-g}
   \biggl\{  {\cal L}\Bigl( g, {\cal R}, \nabla_{a_1}{\cal R}, \ldots,
   \nabla_{{\scriptscriptstyle
(}a_1}\!\!\cdot\!\cdot\!\cdot\!\nabla_{a_{m-1}{\scriptscriptstyle )}}{\cal
R},
   V^m_{a_1\cdots a_m},
   {\hbox{$\psi$'s}} \Bigr) \nonumber\\
& &\qquad\qquad\qquad\qquad\quad{\ }
   -\Bigl[ (\nabla_{a_m} U_m^{a_1\cdots a_m})
   \nabla_{{\scriptscriptstyle
(}a_1}\!\!\cdot\!\cdot\!\cdot\!\nabla_{a_{m-1}{\scriptscriptstyle )}}{\cal
R}
   \,+\, U_m^{a_1\cdots a_m} V^m_{a_1\cdots a_m} \Bigr]
    \biggr\} \ ,\eqnum{3.4}
\end{eqnarray}
in which the highest derivative of the Riemann tensor
is the $(m-1)^{\rm th}$ derivative.

The action (3.4) yields the same equations of motion as the
original action (3.1), but (3.4) is not
entirely equivalent to (3.1) because boundary terms were discarded
in its derivation. The change of boundary terms implies a change
of boundary conditions for the variational problem. However, the
boundary terms at $\partial{\cal M}$ that should be present in the final
Hamiltonian form (2.2) of the action are known: there should be no
boundary terms at $\partial{\cal M}$. Thus, we are free to discard any
boundary terms that arise through integration by parts in spacetime.
At the end of the analysis, any remaining boundary terms at
$\partial{\cal M}$ must be eliminated {}from the Hamiltonian form of the
action anyway.

The algorithm described above can be iterated
until all derivatives of the Riemann tensor have been eliminated.
In the process, sets of auxiliary variables $U_{m-1}$, $V^{m-1}$,
$\ldots$, $U_1$, $V^1$ are introduced, which serve to eliminate
the $(m-1)^{\rm th}$, $\ldots$, $(1)^{\rm th}$ derivatives of
${\cal R}_{bcde}$. The resulting action is
\begin{eqnarray}
& & {\cal S}[g,\psi,U_m,V^m,\ldots,U_1,V^1] \nonumber\\
& &\qquad = i\int_{{\cal M}} d^Dx\sqrt{-g} \biggl\{
   {\cal L}(g,{\cal R},V^1,\ldots,V^m,{\hbox{$\psi$'s}}) \nonumber\\
& &\qquad\qquad\qquad\qquad\quad{\ } -\Bigl[ (\nabla_{a_1}U_1^{a_1}) {\cal
R}
    + U_1^{a_1} V^1_{a_1} \Bigr]  -
    \Bigl[ (\nabla_{a_2}U_2^{a_1a_2})V^1_{a_1} +
    U_2^{a_1a_2} V^2_{a_1a_2} \Bigr] - \cdots \nonumber\\
& &\qquad\qquad\qquad\qquad\quad{\ }\cdots -
    \Bigl[ (\nabla_{a_m}U_m^{a_1\cdots a_m})
    V^{m-1}_{a_1\cdots a_{m-1}}
    + U_m^{a_1\cdots a_m} V^m_{a_1\cdots a_m} \Bigr] \biggr\}
\ .\eqnum{3.5}
\end{eqnarray}
Now isolate the Riemann tensor by introducing one more set of
auxiliary fields, $U_0$ and $V^0$. (This is one more iteration of
the algorithm, but without the integration by parts). The action
becomes
\begin{eqnarray}
& &{\cal S}[g,\psi,U_m,V^m,\ldots,U_0,V^0] \nonumber\\
& &\qquad = i\int_{{\cal M}} d^Dx \sqrt{-g} \biggl\{
   {\cal L}(g,V^0,\ldots,V^m,{\hbox{$\psi$'s}})  \nonumber\\
& &\qquad\qquad\qquad\qquad\quad{\ }  +\Bigl[ U_0 {\cal R} -  U_0 V^0
\Bigr]
    -\Bigl[ (\nabla_{a_1}U_1^{a_1}) V^0
    + U_1^{a_1} V^1_{a_1} \Bigr] -\cdots \nonumber\\
& &\qquad\qquad\qquad\qquad\quad{\ }\cdots - \Bigl[
    (\nabla_{a_m}U_m^{a_1\cdots a_m}) V^{m-1}_{a_1\cdots a_{m-1}}
    + U_m^{a_1\cdots a_m} V^m_{a_1\cdots a_m} \Bigr] \biggr\}
\ .\eqnum{3.6}
\end{eqnarray}
Observe that when the $V$ equations of motion hold,
\begin{eqnarray}
0 & = & -U_0 - \nabla_{a_1} U_1^{a_1} +
        \frac{\partial{\cal L}}{\partial V^0} \ ,\nonumber\\
&\vdots &\nonumber\\
0 & = & - U_{m-1}^{a_1\cdots a_{m-1}} - \nabla_{a_m} U_m^{a_1\cdots a_m}
    + \frac{\partial{\cal L}}{\partial V^{m-1}_{a_1\cdots a_{m-1}} }
    \ ,\nonumber\\
0 & = & - U_{m}^{a_1\cdots a_m} +
    \frac{\partial{\cal L}}{\partial V^m_{a_1\cdots a_m}} \ ,\nonumber
\end{eqnarray}
and the $U$ equations of motion hold,
\begin{eqnarray}
0 & = & {\cal R} - V^0 \ ,\nonumber\\
0 & = & \nabla_{a_1} V^0 - V^1_{a_1}  \ ,\nonumber\\
&\vdots &\nonumber\\
0 & = & \nabla_{{\scriptscriptstyle (}a_m} V^{m-1}_{a_1\cdots
a_{m-1}{\scriptscriptstyle )}}
    - V^m_{a_1\cdots a_m} \ ,\nonumber
\end{eqnarray}
the variable $U_0$ equals
\begin{eqnarray}
U_0^{bcde} & = & \frac{\partial{\cal L}}{\partial({\cal R}_{bcde})}
   - \nabla_{a_1}\biggl(
   \frac{\partial{\cal L}}{\partial (\nabla_{a_1} {\cal R}_{bcde})}
\biggr)
   + \cdots \nonumber\\
   & &\qquad\qquad \cdots + (-1)^m
\nabla_{a_1}\!\!\cdot\!\cdot\!\cdot\!\nabla_{a_m}
   \biggl( \frac{\partial{\cal L}}{\partial
   (\nabla_{{\scriptscriptstyle
(}a_1}\!\!\cdot\!\cdot\!\cdot\!\nabla_{a_m{\scriptscriptstyle )}} {\cal
R}_{bcde}) }
   \biggr) \ .\eqnum{3.7}
\end{eqnarray}
This is the variational derivative \cite{Donder} of the Lagrangian
${\cal L}$ with respect to the Riemann tensor ${\cal
R}_{bcde}$.\footnote{It is
possible to carry out the analysis above
without the fields $V$. For example, in order to eliminate the
$m^{\rm th}$ derivative of the Riemann tensor {}from the action (3.1),
one can perform a Legendre transformation in which
$\nabla_{{\scriptscriptstyle
(}a_m}(\nabla_{a_1}\!\!\cdot\!\cdot\!\cdot\!\nabla_{a_{m-1}{\scriptscriptstyle
)}}{\cal R})$
play the role of velocities and $U_m^{a_1\cdots a_m}$
play the role of momenta, and follow this with an integration by parts.
(Equivalently,  $V^m$ can be eliminated {}from the action (3.4) by
substitution of the solution of the $V^m$ equation of motion.) One
must allow for the possibility that the relationship between the
velocities and momenta is not invertible, signaling the presence of
constraints. The key equation (3.7) and the form (3.8) of the action
(see below) can be deduced in this way, in spite of the fact that the
constraints are not known explicitly.}
\subsection{Elimination of higher--order derivatives of the matter fields}
The second step in the derivation of the Hamiltonian form of the action
is the elimination of all but the first covariant derivatives of the
matter
fields. This can be achieved by applying the same algorithm
that was used in the elimination of derivatives of the Riemann
tensor. The resulting action depends on the first covariant derivative
of the following fields: $U_1,\ldots,U_m$, some (or all) of the matter
fields
$\psi$, and some of the auxiliary fields that were introduced in the
elimination of the higher--order derivatives of $\psi$. I will denote
these fields collectively by $\Psi'$. The covariant derivatives
$\nabla_a\Psi'$ can be isolated as linear terms in the Lagrangian through
the introduction
of yet another set of auxiliary fields, just as the Riemann tensor was
isolated by the introduction of the fields $U_0$, $V^0$.
The action now takes the form
\begin{equation}
{\cal S}[g,\Psi] = i\int_{{\cal M}} d^Dx\sqrt{-g} \Bigl\{ U_0^{abcd} {\cal
R}_{abcd}
+ f(g,\Psi,\nabla_a\Psi') \Bigr\} \ ,\eqnum{3.8}
\end{equation}
where $\nabla_a\Psi'$ appear linearly in $f$ with
coefficients that are independent variables.
In Eq.~(3.8), $\Psi$ denotes the original matter fields $\psi$, the
auxiliary fields $U$ and $V$, and the auxiliary fields that were
introduced in the elimination of the higher--order derivatives
of $\psi$ and the isolation of the first derivatives $\nabla_a\Psi'$.
Thus, $\Psi'$ is the subset of fields $\Psi$ that appear differentiated
in the action. The presentation below is simplified if we assume
that the fields $\Psi'$ are covariant in their tensor indices. (As
discussed at the beginning of this section, some matter fields might
also carry internal indices, such as a Yang--Mills index or a tetrad
index.) This assumption entails no loss of generality---for each
field with a contravariant tensor index that appears
differentiated in the action, say $\psi^a$, a simple change of variables
$\psi_a = g_{ab}\psi^b$, $g_{ab} = g_{ab}$ allows us to replace $\psi^a$
with $\psi_a$ as the fundamental variable.
\subsection{Space--time decomposition}
The third step in the derivation of the Hamiltonian form of the action
is the introduction of a space--time split. Let spacetime have topology
${\cal M}= \Sigma\times I$ and let $t$ label
the hypersurfaces of the foliation $\Sigma$.
The unit normal of the hypersurfaces
is $u_a = -N\nabla_a t$, where $N= [-(\nabla_a t)g^{ab}(\nabla_b
t)]^{-1/2}$
defines the lapse function. The hypersurface metric is defined by
$h_{ab} = g_{ab} + u_a u_b$, so the spacetime metric $g_{ab}$ becomes
\begin{equation}
g_{ab} = h_{ab} - u_au_b \ .\eqnum{3.9}
\end{equation}
The Gauss, Codazzi, and Ricci equations imply (see, for example,
Ref.~\cite{York})
\begin{eqnarray}
{\cal R}_{abcd} & = & R_{abcd} + 2 K_{a{\scriptscriptstyle [}c}
K_{d{\scriptscriptstyle ]}b}
   + 4(D_{{\scriptscriptstyle [}a} K_{b{\scriptscriptstyle
][}c})u_{d{\scriptscriptstyle ]}}
   + 4(D_{{\scriptscriptstyle [}c} K_{d{\scriptscriptstyle
][}a})u_{b{\scriptscriptstyle ]}} \nonumber\\
   & & - 4 u_{{\scriptscriptstyle [}a} \Bigl( {\mbox{$\pounds$}}_u
K_{b{\scriptscriptstyle ][}c} +
        K_{b{\scriptscriptstyle ]}}^e K_{e{\scriptscriptstyle [}c}
        + (D_{b{\scriptscriptstyle ]}} D_{{\scriptscriptstyle [}c} N)/N
\Bigr) u_{d{\scriptscriptstyle ]}}
\ ,\eqnum{3.10}
\end{eqnarray}
where ${\mbox{$\pounds$}}_u$ is the Lie derivative along $u^a$ and
$R_{abcd}$, $K_{ab} = -\frac{1}{2}{\mbox{$\pounds$}}_u h_{ab}$, and
$D_a$ denote the Riemann tensor, extrinsic curvature, and
covariant derivative of the $t={\rm const}$ hypersurfaces,
respectively.

Under the space--time decomposition the tensor fields
$\Psi$ are projected normally and tangentially to the $t={\rm const}$
hypersurfaces. For example, $\psi_a$ is
split into its projections $u^a \psi_a$ and $h_a^b\psi_b$ and $\psi^a$
is split into $u_a \psi^a$ and $h^a_b\psi^b$. The
first derivative $\nabla_a\Psi'$ of a covariant tensor $\Psi'$
is split into hypersurface covariant derivatives and normal Lie
derivatives
of the projections of $\Psi'$. For example, for a scalar
field $\psi$ we have $\nabla_a\psi = - u_a{\mbox{$\pounds$}}_u\psi +
D_a\psi$ and
for a covariant vector field $\psi_a$ we have
\begin{eqnarray}
\nabla_a\psi_b & = & u_au_b{\mbox{$\pounds$}}_u(u^c\psi_c) -
u_a{\mbox{$\pounds$}}_u(h_b^c\psi_c) + D_a(h_b^c\psi_c) - u_b
D_a(u^c\psi_c)
\nonumber\\
& & + K_{ab}u^c\psi_c - 2u_{{\scriptscriptstyle
(}a}K^c_{b{\scriptscriptstyle )}}\psi_c
-u_au_b \psi_c (D^cN)/N + u_a u^c\psi_c (D_bN)/N \ .\eqnum{3.11}
\end{eqnarray}
Note that the Lie derivatives ${\mbox{$\pounds$}}_u$ and the spatial
covariant
derivative of the lapse function, in the combination $(D_aN)/N$, appear
linearly in $\nabla_a\psi_b$. Also note that
${\mbox{$\pounds$}}_u(h_b^c\psi_c)$ is a
spatial tensor; that is, $u^b{\mbox{$\pounds$}}_u(h_b^c\psi_c) = 0$.
The decomposition for the derivatives of higher rank covariant
tensors is similar to that in Eq.~(3.11). I will use ${\cal P}\Psi$ as a
shorthand notation for the normal and tangential projections of the
fields $\Psi$.

With the space--time split described above, the action (3.8) becomes
\begin{eqnarray}
& &{\cal S} = i\int_{{\cal M}} d^Dx \sqrt{-g} \biggl\{ U_0^{abcd}\Bigl[
R_{abcd}
+ 2K_{ac}K_{bd}\Bigr] +8U_0^{abcd}u_d(D_a K_{bc}) \nonumber\\
& &\qquad\qquad\qquad\qquad\quad -4 U_0^{abcd} u_au_d \Bigl[
{\mbox{$\pounds$}}_u K_{bc} + K_b^eK_{ec} + (D_bD_cN)/N \Bigr] \nonumber\\
& &\qquad\qquad\qquad\qquad\quad +f\bigl( h_{ab},h^{ab},{\cal P}\Psi,
{\mbox{$\pounds$}}_u({\cal P}\Psi'),D_a({\cal P}\Psi'),K_{ab},(D_aN)/N
\bigr)
\biggr\} \ ,\eqnum{3.12}
\end{eqnarray}
where ${\mbox{$\pounds$}}_u({\cal P}\Psi)$ and $(D_aN)/N$ appear linearly
in $f$.
The action (3.12) contains a second time derivative
in the term ${\mbox{$\pounds$}}_uK_{bc} =
-\frac{1}{2}{\mbox{$\pounds$}}_u{\mbox{$\pounds$}}_uh_{ab}$.
This can be removed by promoting the extrinsic curvature
$K_{bc}$ to an independent variable. Thus, introduce an auxiliary
variable $P^{ab}$ and write the action as
\begin{eqnarray}
& &{\cal S} = i\int_{{\cal M}} d^Dx \biggl\{ N P^{ab} \Bigl[
{\mbox{$\pounds$}}_u h_{ab} + 2 K_{ab} \Bigr] \nonumber\\
& &\qquad\qquad\qquad{\ }
+  \sqrt{-g}\, U_0^{abcd}\Bigl[ R_{abcd}
+ 2K_{ac}K_{bd}\Bigr] +8 \sqrt{-g}\, U_0^{abcd}u_d(D_a K_{bc}) \nonumber\\
& &\qquad\qquad\qquad{\ }  -4  \sqrt{-g}\, U_0^{abcd} u_au_d \Bigl[
{\mbox{$\pounds$}}_u K_{bc} + K_b^eK_{ec} + (D_bD_cN)/N \Bigr] \nonumber\\
& &\qquad\qquad\qquad{\ }  + \sqrt{-g}\, f\bigl( h_{ab},h^{ab},{\cal
P}\Psi,
{\mbox{$\pounds$}}_u({\cal P}\Psi'),D_a({\cal P}\Psi'),K_{ab},(D_aN)/N
\bigr)
\biggr\} \ .\eqnum{3.13}
\end{eqnarray}
The action (3.12) is recovered when the solution of the
$P^{ab}$, $K_{ab}$ equations of motion is substituted into the action
(3.13).

Now choose a time flow vector field $t^a$ such that $t^a\nabla_a t = 1$,
and define the shift vector by $V^a = h^a_b t^b$ (not to be confused
with the auxiliary fields $V^0$, $\ldots$, $V^m$).
The Lie derivative ${\mbox{$\pounds$}}_u$, acting on
$h_{ab}$, $K_{ab}$, and the covariant tensors ${\cal P}\Psi'$, is
expressed as $N{\mbox{$\pounds$}}_u = {\mbox{$\pounds$}}_t -
{\mbox{$\pounds$}}_V$. With the fields mapped
{}from ${\cal M}$ to $\Sigma\times I$, the Lie derivatives
${\mbox{$\pounds$}}_t$ along
$t^a$ become ordinary time derivatives (denoted by a dot) and
$\sqrt{-g} = N\sqrt{h}$ where $h$ is the determinant of the metric
$h_{ij}$ on $\Sigma$. The action (3.13) becomes
\begin{eqnarray}
& &{\cal S}[N,V,h,K,P,{\cal P}\Psi] \nonumber\\
& &\quad = i\int dt\int_{\Sigma} d^dx \biggl\{ P^{ij} \Bigl[
{\dot h}_{ij} - D_{{\scriptscriptstyle (}i}V_{j{\scriptscriptstyle )}} +
2NK_{ij} \Bigr] \nonumber\\
& &\qquad\qquad\qquad\qquad
+  N\sqrt{h}\, U_0^{ijk\ell}\Bigl[ R_{ijk\ell}
+ 2K_{ik}K_{j\ell}\Bigr] -8 N\sqrt{h}\, U_0^{ijk{\rm u}}(D_i K_{jk})
\nonumber\\
& &\qquad\qquad\qquad\qquad -4  \sqrt{h}\, U_0^{{\rm u}ij{\rm u}} \Bigl[
{\dot K}_{ij} - {\mbox{$\pounds$}}_VK_{ij} + NK_i^kK_{kj} + (D_iD_jN)
\Bigr] \nonumber\\
& &\qquad\qquad\qquad\qquad + N\sqrt{h}\, f\bigl(h_{ij},h^{ij},{\cal
P}\Psi,
{\mbox{$\pounds$}}_u({\cal P}\Psi'),D_i({\cal P}\Psi'),K_{ij},(D_iN)/N
\bigr)
\biggr\} \ ,\eqnum{3.14}
\end{eqnarray}
where $N{\mbox{$\pounds$}}_u({\cal P}\Psi') = ({\cal
P}\Psi')^{\displaystyle\cdot}-{\mbox{$\pounds$}}_V({\cal P}\Psi')$
and $i$, $j$, $\ldots$ are indices for tensors on $\Sigma$. Also,
the notation $U_0^{abc{\rm u}} = -U_0^{abcd}u_d$ and
$U_0^{{\rm u}bc{\rm u}} = U_0^{abcd}u_au_d$ has been used.
\subsection{Hamiltonian form of the action}
The action (3.14) is linear in time derivatives and Lagrange multipliers.
That is, each term in the Lagrangian of Eq.~(3.14) depends linearly on
either the time derivative of a field, the lapse function $N$, or the
shift
vector $V^i$. (Recall that ${\mbox{$\pounds$}}_u({\cal P}\Psi')$ and
$(D_aN)/N$ appear linearly
in $f$.) Therefore each term in the Lagrangian transforms as a
scalar density under reparametrizations in $t$, and no terms such as
$V^i h_{ij} V^j/N$ appear. Moreover, the coefficients of ${\dot h}_{ij}$,
${\dot K}_{ij}$, and $({\cal P}\Psi')^{\displaystyle\cdot}$ are
independent
variables. Specifically, the coefficient of ${\dot h}_{ij}$ is $P^{ij}$,
the coefficient of ${\dot K}_{ij}$ is $-4\sqrt{h}U_0^{{\rm u}ij{\rm u}}$,
and the coefficient of $({\cal P}\Psi')^{\displaystyle\cdot}$ is given
in terms of the auxiliary fields that were introduced in the process of
isolating $\nabla_a\Psi'$ as a linear factor in the action.
It follows that these coefficients (denoted $p_{\alpha}$) are the
canonical momenta conjugate to the coordinates $h_{ij}$,
$K_{ij}$, and ${\cal P}\Psi'$ (denoted $q^{\alpha}$). After all, consider
what happens if one tries to identify both $q^{\alpha}$ and
$p_{\alpha}$ as coordinates, and to define conjugate momenta
$\Pi_{\alpha}^q$ and $\Pi^{\alpha}_p$.
The definition of these momentum variables leads to
sets of second class constraints, namely, $\Pi_{\alpha}^q = p_{\alpha}$
and $\Pi^{\alpha}_p = 0$. Elimination of these constraints through the
Dirac bracket effectively eliminates the new momenta, and reveals the
interpretation of $q^{\alpha}$ and $p_{\alpha}$ as canonically
conjugate variables.

Now remove the spatial derivatives {}from the lapse function $N$ and
the shift vector $V^i$ through integrations by parts.
The action (3.14) takes the ``almost Hamiltonian" form
\begin{eqnarray}
{\cal S}[N,V,q,p,\chi] & = & i\int dt \int_{\Sigma} d^dx \Bigl(
p_{\alpha} {\dot q}^{\alpha} - N{\cal C}_\perp(q,p,\chi)
- V^i{\cal C}_i(q,p) \Bigr) \nonumber\\
& & + i\int dt \int_{\partial\Sigma} d^{d-1}x \sqrt{\sigma}
\Bigl( -4 n_iU_0^{{\rm u}ij{\rm u}} D_jN +
({\hbox{terms $\sim N$ and $V^i$}})\Bigr)
\ ,\eqnum{3.15}
\end{eqnarray}
where $n^i$ is the outward pointing unit normal of the boundary of
space, $\partial\Sigma$, and $\sigma$ is the determinant of the metric
on $\partial\Sigma$. The ``terms $\sim N$ and $V^i$" that appear at
the boundary $\partial\Sigma$ are proportional to the undifferentiated
lapse function $N$ and undifferentiated shift vector $V^i$. Observe that
the action for any diffeomorphism invariant theory can be put into the
form (3.15), since no special assumptions were made in its
derivation.

The action (3.15) is not quite in Hamiltonian form because it
contains certain extra undifferentiated variables $\chi$ in addition to
the canonical variables $q^{\alpha}$ and $p_{\alpha}$ and
Lagrange multipliers $N$ and $V^i$.
These variables include, for example, the
normal and tangential projections of the auxiliary fields $V^0$, $\ldots$,
$V^m$. Note that the $\chi$'s appear only in the Hamiltonian constraint
${\cal C}_\perp$. They do not appear in the momentum constraint ${\cal
C}_i$, or in
the boundary terms proportional to $V^i$, because ${\cal C}_i$ and the
boundary terms proportional to $V^i$ all
originate {}from the Lie derivatives ${\mbox{$\pounds$}}_u$ in the
combination
$p_\alpha ({\dot q}^\alpha - {\mbox{$\pounds$}}_V q^\alpha)$. The $\chi$'s
also do
not appear in the boundary terms proportional to $N$. To see this, one
should recall that these boundary
terms arise through integration by parts that eliminate spatial
derivatives of $N$. Inspection of Eq.~(3.10)  shows
that the spatial derivatives of $N$ appear in the combination
$N{\mbox{$\pounds$}}_u K_{bc} + D_bD_cN$. Thus, the coefficient of
$D_iD_jN$
in the action equals the momentum conjugate to $K_{ij}$, and involves
no $\chi$'s. Likewise, inspection of
Eq.~(3.11) shows that the spatial derivatives of $N$ appear in combination
with Lie derivatives in such a way that the coefficients of $D_iN$
involve the momenta conjugate to ${\cal P}\Psi'$ and ${\cal P}\Psi'$
itself, but
no $\chi$'s.

A true Hamiltonian form of the action is obtained {}from Eq.~(3.15)
by elimination of the variables $\chi$ through
the solution of their algebraic equations of motion,
\begin{equation}
\frac{i}{N} \frac{\delta{\cal S}}{\delta\chi} =
\frac{\partial{\cal C}_\perp}{\partial\chi} = 0 \ .\eqnum{3.16}
\end{equation}
That is, one solves the set of equations (3.16)
and inserts the solution back into the action. If the matrix of
second derivatives of ${\cal C}_\perp$ with respect to the $\chi$'s
has vanishing determinant, then the equations
(3.16) are not independent and cannot be solved
for all of the $\chi$'s as functions of the canonical variables
$q^{\alpha}$ and $p_{\alpha}$. In that case the equations
(3.16) include constraints on the canonical variables.
In any case, when the conditions (3.16) are imposed
the Hamiltonian constraint ${\cal C}_\perp$ is independent of $\chi$.

The constraints that arise through the elimination of the
$\chi$'s must be incorporated into the action principle via Lagrange
multipliers. Let ${\cal C}_{\scriptscriptstyle A}$ denote the
complete set of constraints for the system---the constraints that arise
through equations (3.16) as well as the Hamiltonian
constraint ${\cal C}_\perp$ and momentum constraints ${\cal C}_i$.
Likewise, let
$\lambda^{\scriptscriptstyle A}$
denote the complete set of Lagrange multipliers associated with
the constraints ${\cal C}_{\scriptscriptstyle A}$, including the lapse
function $N$ and shift vector $V^i$. The action, which is now in
Hamiltonian form, reads
\begin{eqnarray}
{\cal S}[\lambda,q,p] & = & i\int dt \int_{\Sigma} d^dx \Bigl(
p_{\alpha} {\dot q}^{\alpha} - \lambda^{\scriptscriptstyle A}{\cal
C}_{\scriptscriptstyle A}(q,p) \Bigr)
\nonumber\\
& & + i\int dt \int_{\partial\Sigma} d^{d-1}x \sqrt{\sigma}
\Bigl( -4 n_iU_0^{{\rm u}ij{\rm u}} D_jN + ({\hbox{terms $\sim N$ and
$V^i$}})\Bigr) \ .\eqnum{3.17}
\end{eqnarray}
This is Eq.~(2.2) with the boundary terms displayed somewhat more
explicitly.

The details of the elimination of the variables $\chi$
must be carried out on a case--by--case basis. The instructive
example of Einstein gravity coupled to Maxwell electrodynamics
is presented in Appendix A.
\subsection{Comments}
Several comments are in order. First, consider the situation in which
one of the constraints, say, ${\cal C}_1$, is simply one of the momentum
variables, say, $p_1$. Then the equation of motion for the Lagrange
multiplier $\lambda_1$ is $p_1 = 0$, and the equation of motion for
$p_1$ yields an expression for $\lambda_1$ in terms of the other
Lagrange multipliers, $q^{\alpha}$, and $p_{\alpha}$. These
equations can be used to eliminate $p_1$ and $\lambda^1$ {}from the
variational principle---effectively one just sets $p_1$ equal to
zero. In this way the pair $q^1$, $p_1$ is removed {}from the list of
dynamical variables in the theory, although in general the action
still depends on $q^1$ (undifferentiated in time). The resulting
situation is
similar to that encountered in the ``almost Hamiltonian" form (3.15)
of the action, in that the action depends on an extra variable. (The
key difference is that the constraints might depend on spatial
derivatives of $q^1$, and also that $q^1$ might appear in the
boundary terms. The variables $\chi$, on the other hand, appeared
only undifferentiated in the Hamiltonian constraint.)
Now one can attempt to eliminate $q^1$ through the solution of
its equation of motion. If the $q^1$ equation of motion can be  solved
for $q^1$, then insertion of this solution into the action yields
a new Hamiltonian form of the action for the system in
which $q^1$ and $p_1$ are completely excluded.
If the $q^1$ equation of motion depends only on the canonical
variables (other than $q^1$ and $p_1$), then $q^1$ is a Lagrange
multiplier and should be left alone. It might happen that the
$q^1$ equation of motion cannot be solved for $q^1$, but also does not
yield a constraint. In this case one can always stick to the
Hamiltonian form of the action that includes $q^1$, $p_1$ and
the constraint ${\cal C}_1 = p_1$.

Although the situation in which one of the constraints is equal to
a momentum variable might appear to be of academic interest only,
it in fact occurs in the examples of Maxwell electrodynamics and
Einstein gravity. In electrodynamics the variable that plays the role
of $q^1$ is the normal projection of the electromagnetic potential, which
becomes a Lagrange multiplier for the Gauss's law constraint.
In Einstein gravity one must first perform a canonical transformation
on the variables $q^\alpha$, $p_\alpha$ to bring the action to
a form in which one of the momentum variables is constrained to
vanish. The variables ($q^1$ and $p_1$) that are eliminated in this
way are the extrinsic curvature $K_{ij}$ and its conjugate.
Details can be found in Appendix A.

As a final comment, observe that the Hamiltonian $H=\int_\Sigma d^dx
\lambda^{\scriptscriptstyle A}{\cal C}_{\scriptscriptstyle A} +
{\hbox{(boundary terms)}}$ obtained {}from
the action functional (3.17) is not necessarily either the total
Hamiltonian or the extended Hamiltonian \cite{HT}.
If the extended Hamiltonian of the system is desired, one
can start with the Hamiltonian $H$ and treat the constraints
${\cal C}_{\scriptscriptstyle A} = 0$ as primary constraints. The
preservation in
time of the primary constraints can lead to secondary
constraints. One then proceeds to the classification
of constraints as first or second class, and to the construction of
the extended Hamiltonian \cite{HT}. For the purpose of this paper,
it is not necessary that the extended Hamiltonian  appear in the
Hamiltonian form (3.17) of the action. What really matters is
that the variational principles based on the action functionals
(3.1) and (3.17) are equivalent.
\section{Black Hole Entropy}
We are now in a position to compute the entropy
of a stationary spacetime with bifurcate Killing horizon
using the microcanonical functional integral method.
It is assumed that the spacetime metric and matter fields
$\{{\tilde g}_{ab}, {\tilde\psi}\}$ satisfy the
classical equations of motion that follow {}from
the action (3.1). Let $t^a$ denote the Killing vector field that
vanishes on the bifurcation surface ${\cal H}$. Consider a spacelike
hypersurface $\Sigma_0$ whose boundaries consist of an
outer boundary ${\cal B}$ and the bifurcation surface ${\cal H}$. Note
that $\Sigma_0$ lies within a single ``wedge" of the spacetime
where $t^a$ is timelike. Now extend $\Sigma_0$ into a foliation
of the wedge by stationary hypersurfaces $t={\rm const}$,
where $t^a\nabla_a t= 1$. Also choose $t^a$ as the time flow vector
field.\footnote{One can choose a time flow vector field
$t^a - \Omega \phi^a$ where $\Omega$ is constant and $\phi^a$
is a spatial Killing vector field, if such a Killing vector field
exists. This changes the shift vector by $-\Omega\phi^a$,
so the argument given in Appendix B that $V^a$ vanishes at ${\cal H}$ no
longer holds. However, the overall results of the analysis are
unchanged because in the action (4.1) below the
extra nonvanishing boundary terms at ${\cal H}$ just cancel
corresponding boundary terms at ${\cal B}$. This can be seen by
reversing the steps that generate the boundary terms
proportional to $V^a$: Sum the identity $p_\alpha {\mbox{$\pounds$}}_V
q^\alpha = 0$
over canonical pairs, integrate over $\Sigma$, then integrate by
parts and use
the momentum constraint. Experience with black hole thermodynamics
in the context of Einstein gravity \cite{BMY,BY3} shows that the
shift vector defined as the spatial projection of $t^a$, not
$t^a -\Omega\phi^a$, is the physically correct definition for the
product of inverse temperature and chemical potential.}
Then the solution $\{{\tilde g}_{ab}, {\tilde\psi}\}$ can
be written in Hamiltonian form $\{{\tilde\lambda},{\tilde q},
{\tilde p}\}$, where $\tilde\lambda$, $\tilde q$, and $\tilde p$
are $t$--independent.

As discussed in Sec.~2, the entropy is obtained by evaluating the
action (2.2) at the complex solution $\{{\bar\lambda},{\bar q},
{\bar p}\}$ with periodic identification in $t$. The complex
solution $\{{\bar\lambda},{\bar q}, {\bar p}\}$ is obtained {}from the
real Lorentzian solution $\{{\tilde\lambda},{\tilde q}, {\tilde p}\}$
by the substitution $t\to -it$. Under reparametrizations in $t$ the
canonical variables transform as scalars and the Lagrange multipliers
transform as scalar densities,  so it follows \cite{Brown} that
$\bar\lambda = -i\tilde\lambda$, $\bar q = \tilde q$, and
$\bar p = \tilde p$. The orbits of the Killing vector field
$t^a$ in the complex spacetime form closed curves (circles) around
the bifurcation surface ${\cal H}$.

According to the discussion of Sec.~2, the action (2.2) contains no
boundary terms at the spacetime boundary
$\partial{\cal M} = {\cal B}\times S^1$ where the boundary data
appropriate
for the microcanonical functional integral are fixed \cite{Brown}.
The boundary terms at ${\cal H}$ are
just the boundary terms displayed in Eq.~(3.17). Thus,
the action takes the form
\begin{eqnarray}
{\cal S}[\lambda,q,p] & = & i\int_{S^1} dt \int_{\Sigma} d^dx \Bigl(
p_{\alpha} {\dot q}^{\alpha} - \lambda^{\scriptscriptstyle A}{\cal
C}_{\scriptscriptstyle A}(q,p) \Bigr)
\nonumber\\
& & + i\int _{S^1} dt \int_{{\cal H}} d^{d-1}x \sqrt{\sigma}
\Bigl( -4 n_iU_0^{{\rm u}ij{\rm u}} D_jN + ({\hbox{terms $\sim N$ and
$V^i$}})\Bigr) \ ,\eqnum{4.1}
\end{eqnarray}
where $\partial\Sigma = {\cal H}\cup{\cal B}$. The boundary terms at
${\cal H}$ should be
understood in terms of a limiting procedure, as discussed in Sec.~2.
That is, the boundary terms are defined by an integral over the boundary
of an excised region that surrounds the bifurcation surface ${\cal H}$,
and
the limit is taken as the excised region shrinks to ${\cal H}$.

In Appendix B it is shown that the
lapse function $\tilde N$ and shift vector $\tilde V^a$, and
hence also $\bar N$ and $\bar V^a$, vanish in the limit as
the bifurcation surface is approached. Thus, all boundary ``terms
$\sim N$ and $V^i$" in the action (4.1) vanish upon evaluation
at the complex solution. This assumes that the coefficients of
the ``terms $\sim N$ and $V^i$", which depend solely on the
canonical variables, are well behaved at the bifurcation surface
${\cal H}$. Now, because $\{{\bar\lambda},{\bar q},
{\bar p}\}$  is a stationary solution of the classical equations
of motion, the $p_\alpha {\dot q}^\alpha$ terms and the constraint
terms vanish in the evaluation of the action. The result is that
the entropy (2.3) becomes
\begin{equation}
{\cal S}_{\scriptscriptstyle BH} \approx {\cal S}[{\bar\lambda},{\bar
q},{\bar p}] =
-4i\int_{S^1} dt \int_{{\cal H}} d^{d-1}x
 \sqrt{\sigma}  n_iU_0^{{\rm u}ij{\rm u}} D_j{\bar N}   \ ,\eqnum{4.2}
\end{equation}
The right--hand side of this expression is evaluated at the
complex solution $\{{\bar\lambda},{\bar q}, {\bar p}\}$. However,
for notational simplicity, the bars have been omitted {}from the $p$'s
and $q$'s in Eq.~(4.2). No ambiguity arises since the canonical
variables for the Lorentzian and complex solutions agree. The bar
is retained on the lapse function since the Lagrange multipliers
for the Lorentzian and complex solutions differ by a factor of $-i$.
In the analysis below I will continue the practice of placing bars
or tildes only over the Lagrange multipliers.

As shown in Appendix B,  the gradient of the lapse
function is related to the surface gravity $\tilde\kappa$ by
$\lim D_i{\tilde N} = -\lim\tilde\kappa n_i$, where the
limit is taken in which the bifurcation surface ${\cal H}$ is approached
along a $t={\rm const}$ hypersurface $\Sigma$. (Note that $n^i$ is
the outward pointing normal of $\partial\Sigma$ at the bifurcation
surface, so $n^i$ points
``radially inward" towards ${\cal H}$.) Thus, we obtain
\begin{equation}
{\cal S}_{\scriptscriptstyle BH} \approx
4\int_{S^1} dt \int_{{\cal H}} d^{d-1}x \sqrt{\sigma}
n_iU_0^{{\rm u}ij{\rm u}} n_j \tilde\kappa
\ .\eqnum{4.3}
\end{equation}
Since the surface gravity of a spacetime with bifurcate Killing horizon
is constant over the horizon \cite{Waldbook}, $\tilde\kappa$ can be
removed {}from the integral over ${\cal H}$. Now, the proper circumference
of the
circular orbits of $t^a$ is
$\int_{S^1} dt\sqrt{-{\bar N}^2 + {\bar V}^i{\bar V}_i}$.\footnote{This
is $\int ds$, where
$ds^2 = -N^2dt^2 + h_{ij} (dx^i +V^i dt)(dx^j +V^j dt)$ with $dx^i=0$,
evaluated at the complex solution.} {}From
Appendix B we have the result $\lim {\tilde V}^i/{\tilde N} = 0$,
so the proper circumference, in the limit as the
bifurcation surface is approached, equals $\int_{S^1} dt {\tilde N}$.
The expression $\tilde\kappa = -\lim n^jD_j{\tilde N}$ then shows
that $\int_{S^1} dt\,\tilde\kappa$
equals the rate of change of circumference with respect to radius
for these orbits. The complex geometry will be smooth at
${\cal H}$, and satisfy the classical equations of motion there, only if
the period in $S^1$ is chosen such that
$\int_{S^1} dt\,\tilde\kappa = 2\pi$. The entropy is then
\begin{equation}
{\cal S}_{\scriptscriptstyle BH} \approx
8\pi \int_{{\cal H}} d^{d-1}x \sqrt{\sigma}
n_iU_0^{{\rm u}ij{\rm u}} n_j \ .\eqnum{4.4}
\end{equation}
The right--hand side of this expression for ${\cal S}_{\scriptscriptstyle
BH}$
depends only on the canonical variables, so it can be evaluated
either at the complex solution $\{{\bar\lambda},{\bar q},
{\bar p}\}$ or at the Lorentzian solution
$\{{\tilde\lambda},{\tilde q}, {\tilde p}\}$.

Recalling the definition
$U_0^{{\rm u}bc{\rm u}} = {\tilde U}_0^{abcd} {\tilde u}_a
{\tilde u}_d$ and using the expression
${\tilde\epsilon}_{ab} = 2 \lim {\tilde u}_{{\scriptscriptstyle [}a}
n_{b{\scriptscriptstyle ]}}$
for the binormal of ${\cal H}$, we have
\begin{equation}
{\cal S}_{\scriptscriptstyle BH} \approx
-2\pi \int_{{\cal H}} d^{d-1}x \sqrt{\sigma}
{\tilde\epsilon}_{ab} {\tilde\epsilon}_{cd} {\tilde U}_0^{abcd}
\ . \eqnum{4.5}
\end{equation}
This is the main result, Eq.~(1.2), for the entropy of a spacetime
with bifurcate Killing horizon. Here, ${U}_0^{abcd}$ is the
variational derivative (3.7) of the Lagrangian with respect to
the Riemann tensor.

\section{Other Path Integral Methods}
\subsection{Hilbert action surface term}
The relationship between the Hilbert action surface term method
(ii) \cite{KOP,BTZ} and the microcanonical functional integral
method (i) can be understood as follows.
Consider first the logical outline of the microcanonical
functional integral method. The action ${\cal S}$ is the
integral of the Lagrangian ${\cal L}$ over the manifold
${\cal M} = {\cal B}\times{\hbox{$I$\kern-3.8pt $R$}}^2$. The integral is
split into two
pieces, an integral over the excised region
${\cal B}\times{\rm disk}$ that contains the bifurcation
surface ${\cal H}$, and an integral over the remainder of the
manifold, ${\cal B}\times{\rm annulus}$. Schematically, we have
\begin{equation}
{\cal S} = \int_{\hbox{$\scriptstyle I$\kern-2.4pt $\scriptstyle
R$}^2}{\cal L} = \int_D{\cal L} + \int_A{\cal L} \ ,\eqnum{5.1}
\end{equation}
where $D$ and $A$ refer to the disk and the annulus,
respectively, and the factor ${\cal B}$ has been suppressed for
notational simplicity. The first integral vanishes in
the limit in which the excised region $D$ shrinks to ${\cal H}$,
under the assumption that ${\cal L}$ is smooth.
The second integral is written in Hamiltonian form, which
yields a volume integral ${\cal S}_{\scriptscriptstyle C}$ [the integral
of $p_\alpha{\dot q}^\alpha - \lambda^{\scriptscriptstyle A}{\cal
C}_{\scriptscriptstyle A}$
in Eq.~(4.1)] and terms at the ``inner" boundary $\partial A_i$
and ``outer" boundary $\partial A_o$ of the annulus. The
terms at the outer boundary $\partial A_o$ are discarded. In
the limit as the disk shrinks to ${\cal H}$, the terms at the
inner boundary $\partial A_i$ become boundary terms
$BT_{{\cal H}}$ at the bifurcation surface [the boundary terms in
Eq.~(4.1)]. Thus, the action becomes
\begin{equation}
{\cal S} \to {\cal S}_{\scriptscriptstyle C} + BT_{{\cal H}} \
.\eqnum{5.2}
\end{equation}
When ${\cal S}$ is evaluated at the complex solution,
the canonical action ${\cal S}_{\scriptscriptstyle C}$ vanishes so
that
\begin{equation}
{\cal S} \to BT_{{\cal H}} \ .\eqnum{5.3}
\end{equation}
Only the term proportional to the gradient of the lapse
function $N$ remains in $BT_{{\cal H}}$ since, as shown in
Appendix B, $N$ and $V^i$ both vanish at the bifurcation surface.

In the Hilbert action surface term
method (ii), the action integral is split according to
\begin{equation}
{\cal S} = \int_{\hbox{$\scriptstyle I$\kern-2.4pt $\scriptstyle
R$}^2}{\cal L} = \int_D{\cal L} + \int_{\partial D}\ell
+ \int_A{\cal L} + \int_{\partial A_i}\ell \ ,\eqnum{5.4}
\end{equation}
where $\int \ell$ is the ``Hilbert action surface term";
that is, $\int \ell$ is the surface term that must be
added to $\int{\cal L}$ such that the boundary conditions
include fixation of the metric on the boundary.
Equation (5.4) is, of course, equivalent to Eq.~(5.1)
since the Hilbert action surface terms at $\partial D$
and $\partial A_i$ cancel one another. As in
the microcanonical functional integral method, the
integral over $D$ vanishes in the limit in which
the excised region shrinks to ${\cal H}$. Also the integral over
$A$ can be written in Hamiltonian form. The resulting
action is
\begin{equation}
{\cal S} \to \int_{\partial D}\ell
+ {\cal S}_{\scriptscriptstyle C} + BT'_{{\cal H}} \ ,\eqnum{5.5}
\end{equation}
where $BT'_{{\cal H}} = BT_{{\cal H}} + \int_{\partial A_i}\ell$. (Again,
the terms at the outer boundary are discarded.)

The integrand in $BT'_{{\cal H}}$ must be linear in the lapse and shift
in order to transform properly under reparametrizations in $t$.
But, in fact, the integrand in $BT'_{{\cal H}}$ cannot depend on spatial
derivatives of $N$ or $V^i$. (Actually, spatial derivatives in
a direction tangent to the boundary are allowed, since these
can be removed through integration by parts.) This can be
understood as follows. The boundary conditions
appropriate for ${\cal S}_{\scriptscriptstyle C} + BT'_{{\cal H}}$ (which
equals $\int_A{\cal L}
+ \int_{\partial A_i}\ell$ plus terms at the outer boundary
$\partial A_o$) include fixation of the induced metric
on $\partial A_i$, by definition of the Hilbert action surface term.
Therefore, with the induced metric on $\partial A_i$ denoted
by $\gamma_{mn}$, we have
\begin{equation}
\delta({\cal S}_{\scriptscriptstyle C} + BT'_{{\cal H}}) =
({\hbox{eom's}}) +
({\hbox{bt's at $\partial A_o$}}) +
\int_{\partial A_i} \pi^{mn}\delta\gamma_{mn}
+ ({\hbox{other bt's at $\partial A_i$}})
\eqnum{5.6}
\end{equation}
for some $\pi^{mn}$. Here, ``eom's" are terms that yield the classical
equations of motion and ``bt's at $\partial A_o$" are boundary terms
at $\partial A_o$. The ``other bt's at $\partial A_i$" are boundary
terms at $\partial A_i$ that involve variations of various matter
fields and auxiliary fields,
but do not involve variations of $\gamma_{mn}$ or
variations of derivatives of $\gamma_{mn}$. The induced metric
$\gamma_{mn}$ on the ($D-1$ dimensional) surface $\partial A_i$
can be split into a lapse, shift, and ($D-2$ dimensional) spatial
metric using the slices $t={\rm const}$ and time flow vector
field $t^a$ induced on $\partial A_i$. If the slices $t={\rm const}$
are orthogonal to $\partial A_i$, so that the unit
normal of the slices lies in $\partial A_i$, then the lapse and shift
components of $\gamma_{mn}$ are just the restrictions of $N$ and
$V^i$ to $\partial A_i$. If the slices $t={\rm const}$ are
not orthogonal to $\partial A_i$, then the lapse and shift components
of $\gamma_{mn}$ are constructed algebraically {}from  $N$ and $V^i$
through simple kinematical boost relations \cite{BLY}. Consequently,
the boundary terms at $\partial A_i$ in Eq.~(5.6) depend on the
variations of $N$ and $V^i$, but not on the variations of their
derivatives. Since ${\cal S}_{\scriptscriptstyle C}$
contains no derivatives of $N$ or $V^i$, Eq.~(5.6) shows that
$BT'_{{\cal H}}$ cannot contain derivatives of $N$ or $V^i$.

When the action of Eq.~(5.5) is evaluated at the complex solution,
the canonical action ${\cal S}_{\scriptscriptstyle C}$ vanishes. The
boundary term
$BT'_{{\cal H}}$ is zero since its integrand is linear in the
undifferentiated lapse and
shift, and the lapse and shift vanish at ${\cal H}$. Therefore
\begin{equation}
{\cal S} \to \int_{\partial D}\ell \ ,\eqnum{5.7}
\end{equation}
which shows that the entropy
${\cal S}_{\scriptscriptstyle BH}\approx {\cal S}({\bar\lambda},{\bar
q},{\bar p})$ equals
the Hilbert action surface term for a small disk surrounding the
bifurcation surface ${\cal H}$, evaluated at the complex solution. In
effect, what has been shown is that the Hilbert action surface
term must include the negative of the particular boundary term
displayed in Eq.~(4.1) that is proportional to the gradient of
the lapse function. The minus sign is compensated by the fact
that the normal $n^i$ of $\partial D$ points away {}from ${\cal H}$.
\subsection{Conical deficit angle}
The starting point for the canonical deficit angle method
(iii) \cite{BTZ,SU,Nelson} for computing black hole entropy
is the (grand canonical) partition function
\begin{equation}
Z(\beta) = \sum_{{\cal M}}\int Dg\,D\psi
\,\exp\Bigl({{\cal S}_\beta[g,\psi]}\Bigr) \ .\eqnum{5.8}
\end{equation}
Here, ${\cal S}_\beta$ is the Hilbert action ${\cal S}_\beta =
\int_{\cal M} {\cal L} + \int_{\partial{\cal M}}\ell$ and the inverse
temperature is defined by the lapse component of
the boundary metric $\gamma_{mn}$ according to
$\beta = i\int dt (\rm{lapse})$. When evaluated at the
complex black hole solution, $\beta$ is the
proper length in the $S^1$ direction as measured
orthogonally to the stationary time slices in $\partial{\cal M}$.
The relationship between $Z[\beta]$ and the density
of states (2.1) is spelled out in detail in Ref.~\cite{BY1}.
For our present purposes it is sufficient to note that
the density of states is a function of the internal energy
$E$, where $E$ is defined by the variation of the Hilbert action
with respect to the lapse component of $\gamma_{mn}$. Thus,
the microcanonical action ${\cal S}$ of Eq.~(2.1) and the Hilbert action
${\cal S}_\beta$ differ by boundary terms that include a term of the form
$-\beta E$. The partition function is then given by the
Laplace transform of the density of states:\footnote{To be
precise, the inverse temperature is a function on the system
boundary ${\cal B}$ and, correspondingly, the energy is a surface
density \cite{BMY,BY1}. Thus, the notation used here should be
viewed as  schematic. On the other hand, $\beta$ and
$E$ can be interpreted as the ``zero mode" parts of the inverse
temperature and energy surface density. The final result (5.15)
for the entropy can be understood in this way.}
\begin{equation}
Z(\beta) = \int dE\, \nu(E) e^{-\beta E} \ .\eqnum{5.9}
\end{equation}
With the relationship $\nu(E) \approx \exp({\cal S}_{\scriptscriptstyle
BH}(E))$,
the leading order approximation to the integral in Eq.~(5.9)
is
\begin{equation}
\ln Z(\beta) \approx {\cal S}_{\scriptscriptstyle BH}(E^*) - \beta E^*
\ ,\eqnum{5.10}
\end{equation}
where $E^*$ is the function of $\beta$ such that $E=E^*(\beta)$
extremizes the exponential:
\begin{equation}
\frac{\partial{\cal S}_{\scriptscriptstyle BH}(E)}{\partial
E}\biggr|_{E^*} = \beta
\ .\eqnum{5.11}
\end{equation}
Equations (5.10) and (5.11) just express $\ln Z(\beta)$ (which is
$-\beta$ times the free energy) as a Legendre
transform of the entropy ${\cal S}_{\scriptscriptstyle BH}(E)$.

{}From the relationships above it is easy to show
that the entropy is given by
\begin{equation}
{\cal S}_{\scriptscriptstyle BH}(E^*) \approx \ln Z(\beta) - \beta
\frac{\partial\ln Z(\beta)}{\partial\beta} \ .\eqnum{5.12}
\end{equation}
The zero--loop approximation to the path integral (5.8) yields
$\ln Z(\beta) \approx {\cal S}_\beta[{\bar g},{\bar\psi}] $,
where ${\bar g}$, $\bar\psi$ is the complex black hole solution
that extremizes ${\cal S}_\beta[g,\psi]$ among configurations whose
proper length (period) in the $S^1$ direction equals $\beta$ at
the boundary $\partial{\cal M}$. Then the entropy can be
written as
\begin{equation}
{\cal S}_{\scriptscriptstyle BH}(E^*) \approx {\cal S}_\beta[{\bar
g},{\bar\psi}]
- \beta \lim \frac{{\cal S}_{\beta'}[{\bar g}',{\bar\psi}']
- {\cal S}_\beta[{\bar g},{\bar\psi}]}{\beta'-\beta} \ ,\eqnum{5.13}
\end{equation}
where the limit is taken in which $\beta'\to\beta$.
It is important to recognize that in taking the derivative
with respect to $\beta$, one does {\it not\/} introduce a
conical singularity in the metric ${\bar g}$
(contrary to claims made in the literature).
Rather, when $\beta$ is varied,
the parameters of the solution ${\bar g}$, $\bar\psi$
(notably the black hole mass parameter) vary in such a way that the
configuration remains a smooth solution of the classical equations
of motion. Thus, in Eq.~(5.13), the extremal configuration
with period $\beta'$ is the smooth solution denoted ${\bar g}'$,
${\bar\psi}'$.

Equation (5.13) can be evaluated with the following
trick. Since the action ${\cal S}_{\beta'}$ is stationary at
${\bar g}'$, ${\bar\psi}'$, one can distort this
configuration without affecting the value of the action
to first order. Thus, replace ${\bar g}'$ with
the metric ${\bar g}^\vee$ and leave ${\bar\psi}'$ alone.
In principle ${\bar g}^\vee$ could
be any metric obtained {}from an infinitesimal variation of
${\bar g}'$. Consider as a particular choice for
${\bar g}^\vee$ the metric whose components are identical
to the components of $\bar g$ in the stationary coordinate
system, but with the period in coordinate time $t$ adjusted
so that the proper length in the $S^1$ direction at $\partial{\cal M}$
is $\beta'$ rather than $\beta$.
Thus, ${\bar g}^\vee$ is a smooth, regular solution of the
classical equations of motion everywhere except at the
bifurcation surface ${\cal H}$. At ${\cal H}$ the regularity condition
$\int_{S^1} dt\, {\tilde \kappa} = 2\pi$ does not hold
for ${\bar g}^\vee$, indicating the presence of a conical
singularity.

With the replacement of ${\bar g}'$ by ${\bar g}^\vee$, the
entropy becomes
\begin{equation}
{\cal S}_{\scriptscriptstyle BH}(E^*) \approx {\cal S}_\beta[{\bar g}] -
\beta
\lim \frac{{\cal S}_{\beta'}[{\bar g}^\vee]
- {\cal S}_\beta[{\bar g}]}{\beta'-\beta} \ .\eqnum{5.14}
\end{equation}
(The dependence on matter fields $\psi$ has been dropped for
notational simplicity.)
Now split the action ${\cal S}_\beta[{\bar g}]$ into an
integral $(\int_D{\cal L})[{\bar g}]$ over the disk plus
``other terms" that consist of an integral over the annulus $A$
and boundary integrals at $\partial{\cal M}$. The integral over $D$
vanishes in the limit as
the disk shrinks to ${\cal H}$, since ${\cal L}$ is smooth. The other
terms contain integrals over $t$ and are proportional to
$\beta$. Likewise, split ${\cal S}_{\beta'}[{\bar g}^\vee]$ into
$(\int_D{\cal L})[{\bar g}^\vee]$ plus ``other terms".
The integral over $D$ does not vanish in this case due
to the conical singularity in ${\bar g}^\vee$. The other
terms are identical to the other terms {}from
${\cal S}_\beta[{\bar g}]$ with the exception that they
are proportional to $\beta'$.  Therefore all ``other terms"
in expression (5.14) cancel, and the entropy becomes
\begin{equation}
{\cal S}_{\scriptscriptstyle BH}(E^*) \approx -\beta \lim
\frac{1}{\beta'-\beta}
\biggl(\int_D{\cal L}\biggr)\biggr|_{{\bar g}^\vee}
\ ,\eqnum{5.15}
\end{equation}
The entropy is thus expressed in terms of a spacetime ${\bar g}^\vee$
with a conical singularity. The limit in Eq.~(5.15) is taken in which
$D$ shrinks to ${\cal H}$ and $\beta'$ approaches $\beta$.

The correctness of the result (5.15) can be verified as follows.
Let ${\cal L}$ be the integrand of Eq.~(3.8).\footnote{Iyer and Wald
\cite{IW2} conjectured that the canonical deficit angle method
(iii) is limited to theories in which the Lagrangian is
a linear function of the Riemann tensor. The results here show that
any theory of the form (3.1) can be treated by the conical deficit
angle method if the action is first put into the form (3.8)
in which the Lagrangian is linear in ${\cal R}_{abcd}$.}
The term proportional to $f$ vanishes in the limit
as the disk shrinks to ${\cal H}$. Likewise, most of the terms in
$U_0^{abcd}{\cal R}_{abcd}$ do not contribute to the entropy---only the
term that contains the curvature in the surface orthogonal to ${\cal H}$
survives as $D$ shrinks to ${\cal H}$. This term captures the curvature
of the conical singularity. Inserting a projection onto the
binormal of ${\cal H}$, we have
\begin{equation}
\biggl(\int_D{\cal L}\biggr)\biggr|_{{\bar g}^\vee}
= \frac{i}{4} \int_{D\times{\cal B}} d^Dx
\biggl( \sqrt{-g} U_0^{efgh} \epsilon_{ef}\epsilon_{gh}
\epsilon^{ab}\epsilon^{cd}{\cal R}_{abcd} \biggr)\biggr|_{{\bar g}^\vee}
\ .\eqnum{5.16}
\end{equation}
The binormal can be written as $\epsilon_{ab}
= 2 u_{{\scriptscriptstyle [}a} n_{b{\scriptscriptstyle ]}}$ where $n_a$
and $u_a$ are
orthogonal to ${\cal H}$ and to each other, and $n^a$ lies in a
$t={\rm const}$ surface. Note that $u_a = -N\nabla_a t$ is
imaginary when evaluated at a complex spacetime, so
$-i {\tilde u}_a$ is the real unit vector with square $+1$
for the metric ${\bar g}^\vee$. The basic interpretation
of the Riemann tensor gives
\begin{equation}
n^a(-iu^b)n^c{{\cal R}_{abc}}^d = \delta n^d/A_D \ ,\eqnum{5.17}
\end{equation}
where $\delta n^d$ is the change in the vector $n^c$ as it is
parallel transported around the perimeter of the disk $D$ (first
in the $n^a$ direction, then in the $-i u^b$ direction) and
$A_D$ is the area of the disk. If the disk has deficit angle
$\alpha$, then $\delta n^d = \alpha(-i u^d)$. Thus, we
find $\epsilon^{ab}\epsilon^{cd}{\cal R}_{abcd} = -4\alpha/A_D$
and
\begin{equation}
\biggl(\int_D{\cal L}\biggr)\biggr|_{{\bar g}^\vee}
= - \int_{D\times{\cal B}} d^{D}x
\biggl( \sqrt{g} \epsilon_{ab}\epsilon_{cd}U_0^{abcd}
\alpha/A_D \biggr)\biggr|_{{\bar g}^\vee}
\ .\eqnum{5.18}
\end{equation}
Now, the deficit angle is given by
$\alpha/(2\pi) = (\beta-\beta')/\beta$, so the entropy {}from
Eq.~(5.15) becomes
\begin{equation}
{\cal S}_{\scriptscriptstyle BH}(E^*) \approx -2\pi\int_{\cal H} d^{D-2}x
\biggl( \sqrt{\sigma} \epsilon_{ab}\epsilon_{cd}U_0^{abcd}
\biggr)\biggr|_{{\bar g}}
\ .\eqnum{5.19}
\end{equation}
In the integrand above, $\epsilon_{ab}\epsilon_{cd}U_0^{abcd}$
is a spacetime scalar and in particular it is invariant under
reparametrizations in $t$. Consequently the integrand can be
evaluated at the Lorentzian black hole solution ${\tilde g}$
rather than $\bar g$. The result agrees with Eq.~(4.5).

\section*{Acknowledgments}
I would like to thank T. Jacobson for helpful communications
and J.W. York for helpful discussions and
comments on the manuscript.
\appendix
\section{Maxwell electrodynamics and Einstein gravity}
Consider Maxwell electrodynamics in $D=4$ spacetime dimensions coupled
to the gravitational field. The electromagnetic field contribution to
the action is
\begin{equation}
{\cal S}_{\scriptscriptstyle M} = -\frac{i}{4\pi}\int_{{\cal M}} d^4x
\sqrt{-g} \,
\nabla_{{\scriptscriptstyle [}a} A_{b{\scriptscriptstyle ]}}  g^{ac}
g^{bd}
\nabla_{{\scriptscriptstyle [}c} A_{d{\scriptscriptstyle ]}} \
.\eqnum{A.1}
\end{equation}
In the form of Eq.~(3.8) the action becomes
\begin{equation}
{\cal S}_{\scriptscriptstyle M} = i\int_{{\cal M}} d^4x \sqrt{-g}
\Bigl\{ \Pi^{ab}(\nabla_a A_b
- \Lambda_{ab}) - (\Lambda_{{\scriptscriptstyle [}ab{\scriptscriptstyle
]}} g^{ac} g^{bd}
\Lambda_{{\scriptscriptstyle [}cd{\scriptscriptstyle ]}})/(4\pi) \Bigr\} \
,\eqnum{A.2}
\end{equation}
where $\nabla_a A_b$ appears linearly. The auxiliary
fields are $\Pi^{ab}$ and $\Lambda_{ab}$. The
``almost Hamiltonian" form [cf.~Eq.~(3.15)] of the action is
\begin{eqnarray}
{\cal S}_{\scriptscriptstyle M} & = & i\int dt \int_\Sigma d^3x \sqrt{h}
\Bigl\{ -\Pi^{{\rm u}{\rm u}} {\dot A}_{{\rm u}}
+ \Pi^{{\rm u}i} {\dot A}_i - N{\cal C}_{\perp}^{\scriptscriptstyle M} -
V^i{\cal C}_i^{\scriptscriptstyle M}
\Bigr\} \nonumber\\
& & + i \int dt \int_{\partial\Sigma} d^2x \sqrt{\sigma} n_i\,
\Bigl\{ NA_{\rm u}\Pi^{{\rm u}i}
- NA^i\Pi^{{\rm u}{\rm u}} - V^jA_j\Pi^{{\rm u}i} \Bigr\}
\ ,\eqnum{A.3}
\end{eqnarray}
where the electromagnetic field contribution to the Hamiltonian
and momentum constraints is
\begin{eqnarray}
{\cal C}_{\perp}^{\scriptscriptstyle M} & = & -D^i(\Pi^{{\rm u}{\rm u}}
A_i)
+ \Pi^{{\rm u}{\rm u}} \Lambda_{{\rm u}{\rm u}}
+ D_i(\Pi^{{\rm u}i}A_{\rm u}) - \Pi^{{\rm u}i}\Bigl(K_i^jA_j
+\Lambda_{{\rm u}i}\Bigr) \nonumber\\
& & +\Pi^{i{\rm u}}\Bigl( D_iA_{\rm u} -
K_i^j A_j - \Lambda_{i{\rm u}} \Bigr) - \Pi^{ij}\Bigl( D_iA_j
- K_{ij}A_{\rm u} - \Lambda_{ij}\Bigr) \nonumber\\
& & + (\Lambda_{{\scriptscriptstyle [}ij{\scriptscriptstyle ]}} h^{ik}
h^{j\ell}
\Lambda_{{\scriptscriptstyle [}k\ell{\scriptscriptstyle ]}})/(4\pi) -
(\Lambda_{{\scriptscriptstyle [}i{\rm u}{\scriptscriptstyle ]}}
h^{ij}\Lambda_{{\scriptscriptstyle [}j{\rm u}{\scriptscriptstyle
]}})/(2\pi) \ ,\eqnum{A.4a}\\
{\cal C}_i^{\scriptscriptstyle M} & = & -\Pi^{{\rm u}{\rm u}} D_iA_{\rm u}
- A_i D_j\Pi^{{\rm u}j} - 2 \Pi^{{\rm u}j}D_{{\scriptscriptstyle
[}j}A_{i{\scriptscriptstyle ]}}
\ .\eqnum{A.4b}
\end{eqnarray}
The action is a functional of the coordinates (the $q^\alpha$'s)
$A_{\rm u}$ and $A_i$, the momenta (the $p_\alpha$'s)
$-\sqrt{h}\Pi^{{\rm u}{\rm u}}$ and
$\sqrt{h}\Pi^{{\rm u}i}$, and the extra variables (the
$\chi$'s) $\Pi^{i{\rm u}}$, $\Pi^{ij}$, $\Lambda_{i{\rm u}}$,
$\Lambda_{{\rm u}i}$, $\Lambda_{ij}$, and $\Lambda_{{\rm u}{\rm u}}$.

The $\chi$'s appear in the action undifferentiated and only in
the function ${\cal C}_{\perp}^{\scriptscriptstyle M}$.
In particular $\Pi^{i{\rm u}}$, $\Pi^{ij}$, $\Lambda_{i{\rm u}}$,
$\Lambda_{{\rm u}i}$ and $\Lambda_{ij}$ appear quadratically in
${\cal C}_{\perp}^{\scriptscriptstyle M}$ and can be eliminated by the
solution
of their algebraic equations of motion. Those equations are
straightforward to derive, and the solution is
\begin{eqnarray}
\Pi^{i{\rm u}} & = & - \Pi^{{\rm u}i} \ ,\eqnum{A.5a}\\
\Pi^{ij} & = &-h^{ik}h^{j\ell} D_{{\scriptscriptstyle [}k}
A_{\ell{\scriptscriptstyle ]}} /(2\pi)
        \ ,\eqnum{A.5b}\\
\Lambda_{i{\rm u}} & = & D_iA_{\rm u} - K_i^jA_j \ ,\eqnum{A.5c}\\
\Lambda_{{\rm u}i} & = & -4\pi h_{ij}\Pi^{{\rm u}j} + D_iA_{\rm u}
        - K_i^jA_j \ ,\eqnum{A.5d}\\
\Lambda_{ij} & = & D_iA_j - K_{ij}A_{\rm u} \ .\eqnum{A.5e}
\end{eqnarray}
Inserting this result into ${\cal C}_{\perp}^{\scriptscriptstyle M}$, one
obtains
\begin{equation}
{\cal C}_{\perp}^{\scriptscriptstyle M} = -D^i(\Pi^{{\rm u}{\rm u}} A_i)
+ \Pi^{{\rm u}{\rm u}} \Lambda_{{\rm u}{\rm u}}
+ A_{\rm u} D_i\Pi^{{\rm u}i}  + 2\pi \Pi^{{\rm u}i}h_{ij}\Pi^{{\rm u}j}
+ D_{{\scriptscriptstyle [}i}A_{j{\scriptscriptstyle ]}}
D^{{\scriptscriptstyle [}i}A^{j{\scriptscriptstyle ]}} /(4\pi)
 \eqnum{A.6}
\end{equation}
for the electromagnetic field contribution to the Hamiltonian constraint.

The variable $\Lambda_{{\rm u}{\rm u}}$ appears in the action (A.3),
(A.4b), (A.6)
as a Lagrange multiplier associated with the constraint
$\Pi^{{\rm u}{\rm u}} = 0$. As discussed in Sec.~3, the variables
$\Lambda_{{\rm u}{\rm u}}$, $\Pi^{{\rm u}{\rm u}}$ can be eliminated
through their equations of motion, which amounts to setting
$\Pi^{{\rm u}{\rm u}}$ equal to zero. The coordinate $A_{\rm u}$
conjugate to $-\sqrt{h}\Pi^{{\rm u}{\rm u}}$ remains as an extra
variable in the action, which now reads
\begin{eqnarray}
{\cal S}_{\scriptscriptstyle M} & = & i\int dt \int_\Sigma d^3x \sqrt{h}
\biggl\{ \Pi^{{\rm u}i} {\dot A}_i - V^i\Bigl[ -A_iD_j\Pi^{{\rm u}j}
+ 2\Pi^{{\rm u}j} D_{{\scriptscriptstyle [}i} A_{j{\scriptscriptstyle ]}}
\Bigr] \nonumber\\
& &\qquad\qquad\qquad\quad -N\Bigl[ A_{\rm u} D_i \Pi^{{\rm u}i}
+ 2\pi \Pi^{{\rm u}i}h_{ij} \Pi^{{\rm u}j}
+ D_{{\scriptscriptstyle [}i}A_{j{\scriptscriptstyle ]}}
D^{{\scriptscriptstyle [}i}A^{j{\scriptscriptstyle ]}} /(4\pi)
\Bigr] \biggr\}\nonumber\\
& & + i \int dt \int_{\partial\Sigma} d^2x \sqrt{\sigma} n_i\,
\Bigl\{ N\Pi^{{\rm u}i}A_{\rm u}
 - V^jA_j\Pi^{{\rm u}i} \Bigr\}  \ ,\eqnum{A.7}
\end{eqnarray}
The equation of motion for $A_{\rm u}$ yields the Gauss's law
constraint $D_i\Pi^{{\rm u}i} = 0$. Thus, $A_{\rm u}$ is a
Lagrange multiplier. Now make the changes of variables
${\cal E}^i = \sqrt{h}\Pi^{{\rm u}i}$ for the momentum conjugate
to $A_i$ and $A_t = A_a t^a = -NA_{\rm u} + V^i A_i$ for the Lagrange
multiplier. The result is
\begin{eqnarray}
{\cal S}_{\scriptscriptstyle M} & = & i\int dt \int_\Sigma d^3x
\biggl\{ {\cal E}^i{\dot A}_i - V^i\Bigl[2{\cal E}^j
D_{{\scriptscriptstyle
[}i}A_{j{\scriptscriptstyle ]}}\Bigr]
+ A_t \Bigl[ D_i{\cal E}^i \Bigr] \nonumber\\
& &\qquad\qquad\qquad-N\Bigl[ 2\pi {\cal E}^i{\cal E}_i/\sqrt{h}
+ \sqrt{h} D_{{\scriptscriptstyle [}i}A_{j{\scriptscriptstyle ]}}
D^{{\scriptscriptstyle [}i}A^{j{\scriptscriptstyle ]}}
/(4\pi)\Bigr]
\biggr\}      \nonumber\\
& & - i \int dt \int_{\partial\Sigma} d^2x \sqrt{\sigma}\,
A_t n_i{\cal E}^i/\sqrt{h}  \ ,\eqnum{A.8}
\end{eqnarray}
the Hamiltonian form of the action for the electromagnetic
field coupled to gravity.

Now consider Einstein gravity in $D=4$ spacetime dimensions,
\begin{equation}
{\cal S}_{\scriptscriptstyle E} = i\int_{{\cal M}} d^4x \sqrt{-g}
g^{ac}g^{bd}{\cal R}_{abcd}
\ ,\eqnum{A.9}
\end{equation}
where Newton's constant equals $1/(16\pi)$.
In the form of Eq.~(3.6),
or equivalently (3.8), the action becomes
\begin{equation}
{\cal S}_{\scriptscriptstyle E} = i\int_{{\cal M}} d^4x \sqrt{-g} \Bigl\{
U_0^{abcd} ({\cal R}_{abcd}
- V^0_{abcd}) + g^{ac} g^{bd} V^0_{abcd} \Bigr\} \ .\eqnum{A.10}
\end{equation}
It is assumed that $U_0^{abcd}$ and $V^0_{abcd}$ have the same
symmetries as ${\cal R}_{abcd}$.
The space--time split leads to the action of Eq.~(3.13), where
the function $f$ is given by
\begin{eqnarray}
f & = & h^{ac}h^{bd}V^0_{abcd} + 2 h^{ab} V^0_{{\rm u}ab{\rm u}}
- U_0^{abcd}h_a^eh_b^fh_c^gh_d^hV^0_{efgh} \nonumber\\
& &
+ 4U_0^{abc{\rm u}}h_a^dh_b^eh_c^fV^0_{def{\rm u}}
-4U_0^{{\rm u}ab{\rm u}}h_a^ch_b^dV^0_{{\rm u}cd{\rm u}} \ .\eqnum{A.11}
\end{eqnarray}
By mapping the fields {}from ${\cal M}$ to $\Sigma\times I$ and
integrating
by parts to remove spatial derivatives {}from the lapse and shift,
one obtains the action in ``almost Hamiltonian" form [cf.~Eq.~(3.15)]:
\begin{eqnarray}
{\cal S}_{\scriptscriptstyle E} & = & i\int dt\int_{\Sigma}d^3x \Bigl\{
P^{ij}{\dot h}_{ij}
+ Q^{ij}{\dot K}_{ij} - N{\cal C}^{\scriptscriptstyle E}_\perp - V^i{\cal
C}^{\scriptscriptstyle E}_i \Bigr\}
\nonumber\\
& & + i\int dt\int_{\partial\Sigma} d^2x (\sqrt{\sigma}/\sqrt{h})
n_i\,\Bigl\{ -2P^{ij}V_j - 2Q^{ij}K_{jk}V^k + Q^{ij}D_jN - ND_jQ^{ij}
\Bigr\} \ .\eqnum{A.12}
\end{eqnarray}
Here, the notation $Q^{ij} = -4\sqrt{h}U_0^{{\rm u}ij{\rm u}}$ has
been introduced, and the gravitational contribution to the Hamiltonian
and momentum constraints is
\begin{eqnarray}
{\cal C}^{\scriptscriptstyle E}_\perp & = & - D_iD_jQ^{ij} - 2P^{ij}K_{ij}
- Q^{ij}K_i^kK_{kj}
- \sqrt{h}U_0^{ijk\ell}(R_{ijk\ell} + 2K_{ik}K_{j\ell} - V^0_{ijk\ell})
\nonumber\\
& & + 4\sqrt{h}U_0^{ijk{\rm u}}(2D_iK_{jk} - V^0_{ijk{\rm u}})
- \sqrt{h} h^{ik}h^{j\ell}V^0_{ijk\ell}
- (Q^{ij} + 2\sqrt{h}h^{ij}) V^0_{{\rm u}ij{\rm u}} \ ,\eqnum{A.13a}\\
{\cal C}^{\scriptscriptstyle E}_i & = & -2D_jP^j_i + Q^{jk}D_iK_{jk} -
2D_j(Q^{jk}K_{ki})
\ .\eqnum{A.13b}
\end{eqnarray}
The action is a functional of the coordinates (the $q^\alpha$'s)
$h_{ij}$ and $K_{ij}$, the momenta (the $p_\alpha$'s) $P^{ij}$ and
$Q^{ij}$, and the extra variables (the $\chi$'s) $U_0^{ijk\ell}$,
$U_0^{ijk{\rm u}}$, $V^0_{ijk\ell}$, $V^0_{ijk{\rm u}}$, and
$V^0_{{\rm u}ij{\rm u}}$.

The $\chi$ variables $U_0^{ijk\ell}$,
$U_0^{ijk{\rm u}}$, $V^0_{ijk\ell}$, and $V^0_{ijk{\rm u}}$ can
be eliminated by the solution of their algebraic equations of motion.
That solution is
\begin{eqnarray}
U_0^{ijk\ell} & = & h^{i{\scriptscriptstyle [}k}h^{\ell{\scriptscriptstyle
]}j} \ ,\eqnum{A.14a}\\
U_0^{ijk{\rm u}} & = & 0 \ ,\eqnum{A.14b}\\
V^0_{ijk\ell} & = & R_{ijk\ell} + 2K_{i{\scriptscriptstyle
[}k}K_{\ell{\scriptscriptstyle ]}j}
\ ,\eqnum{A.14c}\\
V^0_{ijk{\rm u}} & = & 2D_{{\scriptscriptstyle [}i}K_{j{\scriptscriptstyle
]}k} \ .\eqnum{A.14d}
\end{eqnarray}
Inserting this result into ${\cal C}^{\scriptscriptstyle E}_\perp$, one
obtains
\begin{eqnarray}
{\cal C}^{\scriptscriptstyle E}_\perp & = & - D_iD_jQ^{ij} - 2P^{ij}K_{ij}
- Q^{ij}K_i^kK_{kj}
\nonumber\\
& & -\sqrt{h}(R + K^2 - K_{ij}K^{ij})
- (Q^{ij} + 2\sqrt{h}h^{ij}) V^0_{{\rm u}ij{\rm u}} \ .\eqnum{A.15}
\end{eqnarray}
Clearly the variable $V^0_{{\rm u}ij{\rm u}}$ plays the
role of a Lagrange multiplier for the constraint
$Q^{ij} + 2\sqrt{h}h^{ij} = 0$. The situation here is close to that
discussed in Sec.~3e in which a constraint (denoted ${\cal C}_1$)
is given by a momentum variable (denoted $p_1$). In fact, the present
theory can be placed in this form by a canonical transformation in which
$Q^{ij}$ is replaced by $Q^{ij} + 2\sqrt{h}h^{ij}$ as the momentum
conjugate to $K_{ij}$. The form of the canonical transformation can be
deduced {}from the relationship
\begin{eqnarray}
& & P^{ij}{\dot h}_{ij} + Q^{ij}{\dot K}_{ij} \nonumber\\
& &\qquad\quad = (P^{ij} + \sqrt{h}Kh^{ij}
- 2\sqrt{h}K^{ij}){\dot h}_{ij} + (Q^{ij} + 2\sqrt{h}h^{ij}){\dot K}_{ij}
-2(\sqrt{h}K)^{\displaystyle\cdot} \ .\eqnum{A.16}
\end{eqnarray}
Thus, define the new momenta
\begin{eqnarray}
{\bar P}^{ij} & = & P^{ij} + \sqrt{h}Kh^{ij} - 2\sqrt{h}K^{ij}
\ ,\eqnum{A.17a}\\
{\bar Q}^{ij} & = & Q^{ij} + 2\sqrt{h}h^{ij} \ ,\eqnum{A.17b}
\end{eqnarray}
and the action becomes
\begin{eqnarray}
{\cal S}_{\scriptscriptstyle E} & = & i\int dt\int_{\Sigma}d^3x \Bigl\{
{\bar P}^{ij}{\dot h}_{ij}
+ {\bar Q}^{ij}{\dot K}_{ij} - 2(\sqrt{h}K)^{\displaystyle\cdot}
- N{\cal C}^{\scriptscriptstyle E}_\perp - V^i{\cal C}^{\scriptscriptstyle
E}_i \Bigr\}
\nonumber\\
& & + i\int dt\int_{\partial\Sigma} d^2x (\sqrt{\sigma}/\sqrt{h})
n_i\,\Bigl\{ -2{\bar P}^{ij}V_j - 2{\bar Q}^{ij}K_{jk}V^k +
{\bar Q}^{ij}D_jN
\nonumber\\
& &\qquad\qquad\qquad\qquad\qquad\qquad\quad
- ND_j{\bar Q}^{ij} - 2\sqrt{h}D^iN + 2\sqrt{h}KV^i
\Bigr\} \ ,\eqnum{A.18}
\end{eqnarray}
where
\begin{eqnarray}
{\cal C}^{\scriptscriptstyle E}_\perp & = & - D_iD_j{\bar Q}^{ij} - 2{\bar
P}^{ij}K_{ij}
- {\bar Q}^{ij}K_i^kK_{kj}
\nonumber\\
& & -\sqrt{h}(R - K^2 + K_{ij}K^{ij})
- {\bar Q}^{ij} V^0_{{\rm u}ij{\rm u}} \ .\eqnum{A.19a}\\
{\cal C}^{\scriptscriptstyle E}_i & = & -2D_j{\bar P}^j_i + {\bar
Q}^{jk}D_iK_{jk}
- 2D_j({\bar Q}^{jk}K_{ki})
\ .\eqnum{A.19b}
\end{eqnarray}
The variables ${\bar Q}^{ij}$ and $V^0_{{\rm u}ij{\rm u}}$ (which
play the role of $p_1$ and $\lambda^1$ in the discussion of Sec.~3e)
can  be eliminated by setting ${\bar Q}^{ij}$ equal to zero. The
action then reduces to
\begin{eqnarray}
{\cal S}_{\scriptscriptstyle E} & = & i\int dt\int_{\Sigma}d^3x \Bigl\{
{\bar P}^{ij}{\dot h}_{ij}
- 2(\sqrt{h}K)^{\displaystyle\cdot} - V^i[-2D_j{\bar P}^j_i] \nonumber\\
& & \qquad\qquad\qquad -N\Bigl[ - 2{\bar P}^{ij}K_{ij}
- \sqrt{h}(R - K^2 + K_{ij}K^{ij}) \Bigr] \Bigr\}
\nonumber\\
& & + i\int dt\int_{\partial\Sigma} d^2x \sqrt{\sigma}
n_i\,\Bigl\{ -2{\bar P}^{ij}V_j/\sqrt{h} - 2 D^iN + 2 KV^i
\Bigr\} \ ,\eqnum{A.20}
\end{eqnarray}
where $K_{ij}$ (which plays the role of $q^1$) appears as an extra
independent variable in addition to the canonical pair $h_{ij}$,
${\bar P}^{ij}$, the lapse function $N$, and the shift vector $V^i$.

The equation of motion for $K_{ij}$ which follows {}from the
action (A.20) is
\begin{equation}
0 = -2{\bar P}^{ij} + 2\sqrt{h}Kh^{ij} - 2\sqrt{h}K^{ij} \ ,\eqnum{A.21}
\end{equation}
and the solution of this equation is
\begin{equation}
\sqrt{h}K_{ij} = -{\bar P}_{ij} + {\bar P} h_{ij}/2 \ .\eqnum{A.22}
\end{equation}
By substituting this result into Eq.~(A.20), we find
the action for Einstein gravity in the Hamiltonian form
\begin{eqnarray}
{\cal S}_{\scriptscriptstyle E} & = & i\int dt\int_\Sigma d^3x \Bigl\{
-{\dot{\bar P}}^{ij} h_{ij} - V^i[-2D_j{\bar P}^j_i]
-N\Bigl[ (2{\bar P}^{ij}{\bar P}_{ij}
- {\bar P}^2)/(2\sqrt{h}) - \sqrt{h}R \Bigr] \Bigr\}\nonumber\\
& & + i\int dt\int_{\partial\Sigma} d^2x \sqrt{\sigma}
n_i\,\Bigl\{ V_j({\bar P}h^{ij} -2{\bar P}^{ij})/\sqrt{h} - 2D^iN
\Bigr\} \ ,\eqnum{A.23}
\end{eqnarray}
Note that the ``kinetic" term of Eq.~(A.23) is
$-{\dot{\bar P}}^{ij}h_{ij}$,
rather than the more usual ${\bar P}^{ij}{\dot h}_{ij}$. The difference
is just a boundary term, $i\int_\Sigma d^3x\,{{\bar P}}$ at the
initial and final times. If we had originally chosen the
action (A.9) to include a boundary term $2i\int_\Sigma d^3x\,\sqrt{h}K$
at the initial and final times, then we would have obtained
${\bar P}^{ij}{\dot h}_{ij}$ for the kinetic term in the Hamiltonian
form of the action.

\section{Surface gravity}
As described in Sec.~4, one wedge of the spacetime ${\tilde g}_{ab}$ is
foliated into stationary hypersurfaces $t={\rm const}$ with time flow
vector field $t^a$,
where $t^a$ is the Killing vector field that vanishes at the bifurcation
surface ${\cal H}$. In this appendix I will drop the tildes with the
understanding that all relationships hold for the Lorentzian metric
${\tilde g}_{ab}$.

The lapse function $N$ and shift vector $V^a$ satisfy
\begin{equation}
t^a = {N}{u}^a + {V}^a \ .\eqnum{B.1}
\end{equation}
This expression is well defined on the interior of the wedge. Consider
the limit in which ${\cal H}$ is approached {}from within a
$t={\rm const}$ hypersurface, say, $\Sigma_0$.
Since $t^a$ vanishes at ${\cal H}$, it follows by contraction of
Eq.~(B.1) successively with $u_a$ and $h_a^b$ that
\begin{equation}
\lim N = 0 \ ,\qquad \lim V^a = 0 \ .\eqnum{B.2}
\end{equation}
This result is used in Sec.~4 to show that the ``terms $\sim N$ and
$V^i$" vanish at the boundary ${\cal H}$ of $\Sigma$.

The surface gravity of a Killing horizon equals \cite{Waldbook}
\begin{equation}
\kappa = \lim (|t||a|) \ ,\eqnum{B.3}
\end{equation}
where $|t| = \sqrt{-t^a t_a}$ is the magnitude of $t^a$ and
$|a|=\sqrt{a^c a_c}$ is the magnitude
of the acceleration $a^c = (t^a\nabla_a t^c)/|t|^2$ of the orbits of
$t^a$.
I will now show that, in effect, the shift vector $V^a$ vanishes
sufficiently rapidly in the limit as the bifurcation surface is approached
{}from within $\Sigma_0$ so that the orbits of $t^a$ become orthogonal to
the $t={\rm const}$ surfaces. Then the surface gravity can be expressed
in terms of the acceleration of the unit normal $u^a$ of the
$t={\rm const}$ surfaces.

Using the Killing vector field property
$\nabla_{{\scriptscriptstyle (}a} t_{b{\scriptscriptstyle )}} = 0$, one
can easily show that
$|t|a_c = \nabla_c|t|$. Then the surface gravity (B.3) becomes
\begin{equation}
\kappa = \lim \sqrt{(\nabla^a|t|)(\nabla_a|t|)} \ .\eqnum{B.4}
\end{equation}
Observe that the limit of
$|t| = \sqrt{-t^a t_a} = \sqrt{N^2 - V^aV_a}$ is a constant
(namely zero) as ${\cal H}$ is approached {}from within $\Sigma_0$. Thus
we have  $\lim \sigma^{ab}\nabla_b|t|= 0$
where $\sigma_{ab} = h_{ab} - n_a n_b$ is the induced metric on
${\cal H}$ and $n_a$ is the unit normal of ${\cal H}$ in $\Sigma_0$.
Since also $t^a\nabla_a|t| = 0$, the gradient of $|t|$ lies entirely
in the $v^a$ direction,
\begin{equation}
\lim \nabla_a|t| = \lim v_a v^b\nabla_b {|t|} \ ,\eqnum{B.5}
\end{equation}
where $v^a$ is the unit vector orthogonal to both
$\sigma_{ab}$ and $t_a$:
\begin{equation}
v^a \sim \lim (N n^a + n^b V_b u^a) \ .\eqnum{B.6}
\end{equation}
Because the surface gravity of a bifurcate Killing horizon
is nonzero \cite{Waldbook}, it follows {}from Eq.~(B.4) that
$\lim \nabla_a {|t|}\neq 0$. Also note that $\nabla_a|t|$ is
spacelike, since it is orthogonal to the timelike vector $t^a$.
Therefore $\lim h_a^b \nabla_b|t|\neq 0$ and, since
$\lim \sigma^{ab}\nabla_b|t|= 0$, we conclude that
$\lim\, n^b\nabla_b|t|\neq 0$.

Now consider the limit of $V^a/|t|$ as ${\cal H}$ is approached {}from
within
$\Sigma_0$. This is an indeterminate
form, $0/0$. We can apply l'H\^{o}pital's rule and differentiate
both numerator and denominator along the normal $n^a$ direction within
$\Sigma_0$. The derivative of the denominator, $n^b\nabla_b{|t|}$, has
a nonzero limit by the argument above. The derivative of the numerator
is
\begin{eqnarray}
n^b\nabla_b V^a & = & n^b\nabla_b(h^a_c t^c) \nonumber\\
& = & n^b t^c \nabla_b h^a_c + n^b h^a_c \nabla_bt^c
\ .\eqnum{B.7}
\end{eqnarray}
Using $h^a_c = \delta^a_c + u^a u_c$, one can write the first
term as
\begin{eqnarray}
n^b t^c \nabla_b h^a_c & = & n^b t^c \nabla_b(u^a u_c) \nonumber\\
& = & N K^{ab} n_b - u^a n^b K_{bc} V^c \ ,\eqnum{B.8}
\end{eqnarray}
where $K_{ab} = - h_a^c\nabla_c u_b$ is the extrinsic curvature
of $\Sigma_0$. The two terms in Eq.~(B.8) vanish in the limit
because the lapse $N$ and shift $V^c$ both vanish in the limit.
Using the relation $h^{ac} = \sigma^{ac} + n^a n^c$ and
the Killing vector field property
$\nabla_{{\scriptscriptstyle (}b} t_{c{\scriptscriptstyle )}} = 0$, one
can write the second
term of Eq.~(B.7) as
\begin{eqnarray}
n^b h^{ac} \nabla_b t_c & = & n^b\sigma^{ac} \nabla_b t_c\nonumber\\
& = & -n^b\sigma^{ac} \nabla_c t_b \nonumber\\
& = & - n^b \sigma^{ac} \nabla_c (N u_b + V_b) \nonumber\\
& = & N\sigma^{ac} K_{cb} n^b - n^b \sigma^{ac}\nabla_c V_b
\ .\eqnum{B.9}
\end{eqnarray}
The first term in Eq.~(B.9) vanishes in the limit due to the
factor of $N$, and the second term in Eq.~(B.9) is zero in the
limit since the derivative acts along the bifurcation surface
where $V_b$ vanishes.
The result is that $\lim n^b\nabla_b V^a = 0$, so by l'H\^{o}pital's
rule we have $\lim V^a/|t| = 0$. Since $|t| = \sqrt{N^2 - V^bV_b}$,
we also find $\lim (V^a/N)(1-V^bV_b/N^2)^{-1/2} = 0$ which implies
$\lim V^a/N = 0$.

The gradient of $|t|$ is given by
\begin{equation}
\nabla_a|t| =  \frac{N \nabla_a N}{|t|}
+ \frac{V^b \nabla_a V_b}{|t|} \ .\eqnum{B.10}
\end{equation}
The results above show that the second term vanishes in the limit as
${\cal H}$ is approached {}from within $\Sigma_0$,
and that $\lim N/|t| = 1$. Therefore Eq.~(B.10) yields
$ \lim \nabla_a|t| = \lim \nabla_a N $. We also find {}from Eq.~(B.6)
that $v^a = n^a$. Thus, Eq.~(B.5) becomes
\begin{equation}
\lim\nabla_a N = \lim n_a n^b\nabla_b N \ .\eqnum{B.11}
\end{equation}
It follows that the surface gravity (B.4) can be written as
$\kappa = \lim |n^a\nabla_a N|$.
If we choose, as in the main body of the paper, the unit normal
$n^a$ to point ``radially inward" towards ${\cal H}$, the surface
gravity becomes $\kappa = -\lim n^a \nabla_a N$. {}From
the relationship (B.11) we find the key result
\begin{equation}
\lim \nabla_a N = -\lim \kappa n_a \ .\eqnum{B.12}
\end{equation}
Finally, note that the surface gravity also can be expressed as
$\kappa = -\lim N n^b A_b$, where (see Ref.~\cite{York})
$A_b = u^a\nabla_a u_b = h_b^c(\nabla_c N)/N$ is the
acceleration of the unit normal of the $t={\rm const}$ hypersurfaces.


\end{document}